\documentclass[useAMS,usenatbib]{mn2e}

\usepackage{lscape,graphicx,amssymb,aas_macros}
\usepackage{amsmath} 
\usepackage{amssymb} 
\usepackage{graphicx} 
\usepackage{indentfirst} 
\usepackage{pdflscape} 
\usepackage{placeins} 
\usepackage{nicefrac} 
\usepackage{afterpage} 
\usepackage{booktabs} 
\usepackage{color} 
\usepackage{float}

\newcommand{\apg}{\:^{>}_{\sim}\:}
\newcommand{\apl}{\:^{<}_{\sim}\:}
\newcommand{\cmjj}{\mbox{${\rm cm^{-2}}$}}
\newcommand{\etal}{et al.}

\newcommand{\kms}{\mbox{km\ s${^{-1}}$}}
\newcommand{\lya}{\mbox{${\rm Ly}\alpha$}}

\title[The Cool ISM\&CGM of Massive Lensing Galaxies at $z=0.4-0.7$]{Probing The Cool Interstellar and Circumgalactic Gas of Three Massive Lensing Galaxies at $z=0.4-0.7$\thanks{Based on data gathered with the 6.5 m Magellan Telescopes located at Las Campanas Observatory, the ESO telescopes at the La Silla Paranal Observatory, and the NASA/ESA Hubble Space Telescope operated by the Space Telescope Science Institute and the Association of Universities for Research in Astronomy, Inc., under NASA contract NAS 5-26555.  Additional data were obtained at the W.M. Keck Observatory, which is operated as a scientific partnership among the California Institute of Technology, the University of California and the National Aeronautics and Space Administration.  The Observatory was made possible by the generous financial support of the W.M. Keck Foundation. }} \author[Zahedy et al.]{Fakhri S. Zahedy$^{1}$\thanks{E-mail:
fsz@uchicago.edu}, Hsiao-Wen Chen$^{1,2}$\thanks{E-mail:
hchen@oddjob.uchicago.edu}, Michael Rauch$^{3}$, Michelle L.\ Wilson$^{4}$, \newauthor and Ann Zabludoff$^{4}$\\
$^{1}$Department of Astronomy \& Astrophysics, The University of Chicago, Chicago, IL 60637, USA \\
$^{2}$Kavli Institute for Cosmological Physics, The University of Chicago, Chicago, IL 60637, USA  \\
$^{3}$The Observatories of the Carnegie Institution for Science, 813 Santa Barbara Street, Pasadena, CA 91101, USA \\
$^{4}$Department of Astronomy, University of Arizona, Steward Observatory, Tucson, AZ 85721, USA}

\begin{document}

\pagerange{\pageref{firstpage}--\pageref{lastpage}} \pubyear{2002}

\maketitle

\label{firstpage}

\begin{abstract}

  We present multi-sightline absorption spectroscopy of cool gas
  around three lensing galaxies at $z=0.4-0.7$.  These lenses have
  half-light radii $r_e=2.6-8$ kpc and stellar masses of
  $\log\,M_*/M_\odot=10.9-11.4$, and therefore resemble nearby passive
  elliptical galaxies.
%  Around 40\% of these massive quiescent
%  galaxies at intermediate redshifts exhibit cool gas at projected
%  distances $d\sim 20-100$ kpc in previous absorption-line surveys.
  The lensed QSO sightlines presented here occur at projected
  distances of $d=3-15$ kpc (or $d\approx 1-2\,r_e$) from the lensing
  galaxies, providing for the first time an opportunity to probe both
  interstellar gas at $r\sim r_e$ and circumgalactic gas at larger
  radii $r\gg r_e$ of these distant quiescent galaxies.  We observe
  distinct gas absorption properties among different lenses and among
  sightlines of individual lenses.  Specifically, while the quadruple
  lens for HE\,0435$-$1223 shows no absorption features to very
  sensitive limits along all four sightlines, strong Mg\,II, Fe\,II,
  Mg\,I, and Ca\,II absorption transitions are detected along both
  sightlines near the double lens for HE\,0047$-$1756, and in one of
  the two sightlines near the double lens for HE\,1104$-$1805.  The
  absorbers are resolved into $8-15$ individual components with a
  line-of-sight velocity spread of $\Delta\,v\approx 300-600$ \kms.
  The large ionic column densities, $\log\,N\apg 14$, observed in two
  components suggest that these may be Lyman limit or damped
  \lya\ absorbers with a significant neutral hydrogen fraction.  The
  majority of the absorbing components exhibit a uniform super solar
  Fe/Mg ratio with a scatter of $<0.1$ dex across the full $\Delta\,v$
  range.  Given a predominantly old stellar population in these
  lensing galaxies, we argue that the observed large velocity width
  and Fe-rich abundance pattern can be explained by SNe~Ia enriched
  gas at radius $r\sim r_e$.
We show that additional spatial constraints in line-of-sight velocity
and relative abundance ratios afforded by a multi-sightline approach
provide a powerful tool to resolve the origin of chemically-enriched
cool gas in massive halos.

\end{abstract}


\section{Introduction}

For over two decades, QSO absorption spectroscopy has provided a
sensitive probe of low-density intergalactic gas, circumgalactic
medium (CGM), and interstellar medium (ISM) which are otherwise too
diffuse to be detected in emission beyond the local universe (e.g.,
Verheijen \etal\ 2007).  Traditionally, the physical properties of the
extended gas around galaxies, such as spatial extent, mean covering
fraction, and total mass content, are characterized using a
statistical approach over an ensemble of projected galaxy--background
QSO pairs.  This approach has yielded statistically significant
constraints for the chemically-enriched CGM around both star-forming
and quiescent galaxies based on searches of strong absorption
transitions, including the Mg\,II\,$\lambda\lambda\,2796, 2803$
absorption doublet (e.g., Bowen \etal\ 1995; Chen \etal\ 2010;
Lovegrove \& Simcoe 2011; Kacprzak \etal\ 2011) and the hydrogen Lyman
series (e.g., Lanzetta \etal\ 1995; Chen \etal\ 1998; Tripp
\etal\ 1998; Rudie \etal\ 2013; Tumlinson \etal\ 2013; Liang \& Chen
2014; Johnson \etal\ 2015).  In contrast, absorption-line observations
of diffuse gas in interstellar space have been limited because the ISM
of distant galaxies has a much smaller cross section and closely
projected QSO and galaxy pairs are rare.

Despite significant progress in characterizing the CGM, the origin of
cool, metal-enriched gas around galaxies remains ambiguous.  Possible
mechanisms to produce cool gas in a hot halo include outflows from
super-galactic winds (e.g., Murray \etal\ 2011; Booth \etal\ 2013),
stripped satellites due to tidal interactions or ram pressure (e.g.,
Wang 1993; Agertz \etal\ 2009; Gauthier 2013), gas accreted from the
intergalactic medium (IGM, e.g., Rauch \etal\ 1997; Nelson
\etal\ 2013), as well as in-situ cloud formation from thermal
instabilities (e.g., Mo \& Miralda-Escude 1996; Maller \& Bullock
2004; Sharma \etal\ 2012).

A number of CGM observations targeting star-forming galaxies have
suggested that one or a combination of the above scenarios are at
play, such as outflows (Bordoloi \etal\ 2011) or a combination of
outflowing and infalling gas (e.g., Bouch{\'e} \etal\ 2012; Kacprzak
\etal\ 2012).  These findings have been based on a simple assumption
that outflows occur along the rotation (minor) axis, while accretion
proceeds along the disk plane (major axis).  However, complications
arise when considering the required energetics to drive the observed
velocity field of the gas (e.g., Gauthier \& Chen 2012) and possible
spin-filament alignment with the cosmic web (e.g., Tempel \etal\ 2013)
which, in the absence of galactic-scale outflows, would also give rise
to an elevated incidence of absorbers along the minor axis of disk
galaxies.

To better discriminate between different scenarios for the origin of
metal-line absorbers, both spatially resolved gas kinematics and
knowledge of galaxy star formation history are necessary.  As shown in
multi-wavelength imaging observations of local starburst galaxies
(e.g., Suchkov \etal\ 1996; Cecil \etal\ 2001), stronger constraints
for gas flows in galactic halos can be obtained from the observed
spatial variations in the velocity field and physical conditions of
the gas.  In a pilot project, Chen \etal\ (2014) also demonstrated
that multi-sightline observations of the CGM using a quadruply-lensed
QSO, coupled with high resolution imaging of the associated galaxies,
enable direct measurements of the velocity gradient and coherence
length of the absorbing gas.  In turn, these measurements provide
critical constraints for distinguishing between different gas flow
models.

\begin{figure*} 
\includegraphics[width=160mm]{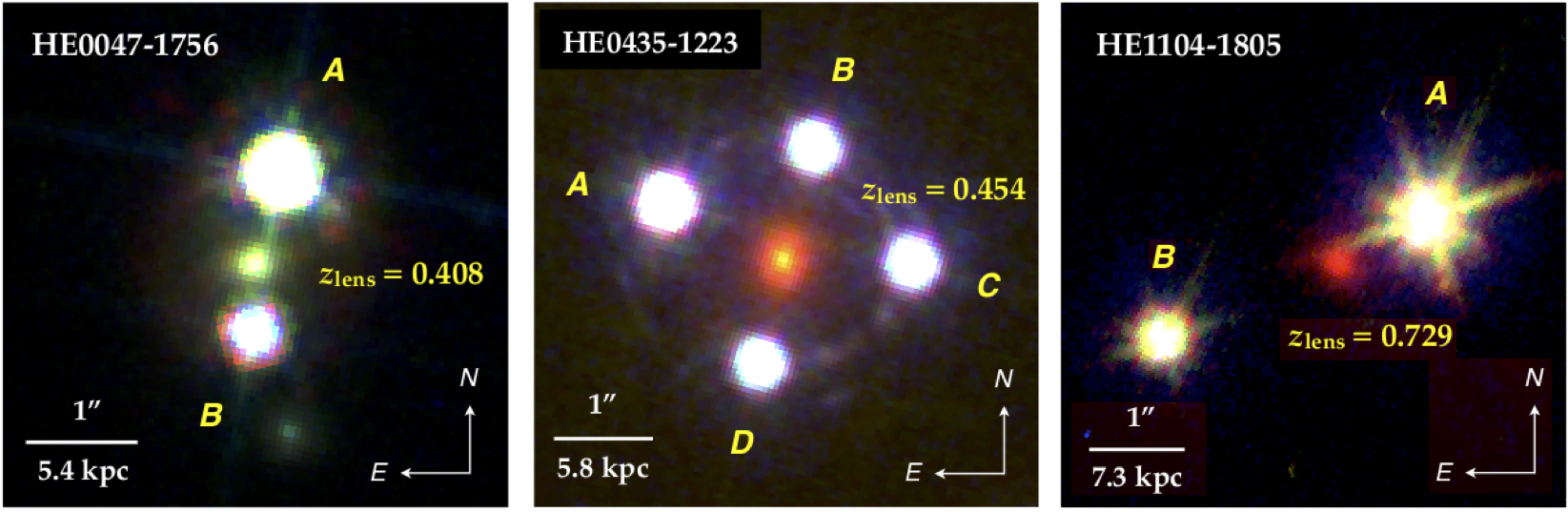}
\caption{False color composite images of three gravitational lens
  systems in our study.  The composite images were produced using the
  HST/ACS WFC and the F555W and F814W filters and NICMOS and the F160W
  filter for HE0047$-$1756, HST/WFC3 UVIS channel and the F275W filter
  and ACS WFC and the F555W and F814W filters for HE\,0435$-$1223, and
  HST/WFC3 UVIS channel and the F275W filter and WFPC2 and the F555W
  and F814W filters for HE\,1104$-$1805 (see Table 1 for details).
  The imaging data were retrieved from the HST data archive (see \S\
  2.1 for descriptions).  For each system, the lensing galaxy is
  located at the center of each panel with the horizontal bar in the
  lower-left corner indicating the $1''$ angular scale and the
  corresponding physical distance at the redshift of the lens.  Using
  absorption spectra of the lensed QSOs, we are able to probe halo gas
  around the lens on projected distance scales of $d \sim 3-15$ kpc
  across the center of each lensing galaxy.  Note that the lensing
  galaxy of HE\,0047$-$1756 exhibit consistently red colors as those
  of HE\,0435$-$1223 and HE\,1104$-$1805 (see Tables 3 \& 4).  The
  available F275W images for the fields around HE\,0435$-$1223 and
  HE\,1104$-$1805 provide strong contraints for a lack of young stars
  in these lensing galaxies.  As a result, the lensing galaxies appear
  to be red in the composite images that include the F275W band.
  However, an UV image is not available for the field around
  HE\,0047$-$1756.  Including available optical and NIR images results
  in the apparent bluer color of this lensing galaxy.}
\label{Figure 1}
\end{figure*}

An added advantage of studying lensed QSO fields is the opportunity of
probing distant ISM by targeting the lensing galaxies.  In this paper,
we apply three multiply-lensed QSOs, HE\,0047$-$1756, HE\,0435$-$1223,
and HE\,1104$-$1805, to study the inner halo gas content of their
lensing galaxies at $z=0.4-0.7$.  By targeting the lensing galaxies,
our study focuses on the cool gas around massive, early-type galaxies
(e.g., Keeton \etal\ 1998).  These galaxies exhibit optical colors and
spectral features that are characteristic of nearby elliptical
galaxies.  Although these evolved galaxies exhibit little/no on-going
star formation, they are not all devoid of cold gas.  Systematic H\,I
and CO searches have uncovered a non-negligible amount of neutral gas
in roughly 40\% of nearby ellipticals (e.g., Oosterloo \etal\ 2010;
Young \etal\ 2014), suggesting that some feedback processes are in
effect to prevent continuing star formation in these gas-rich
quiescent galaxies (e.g., McNamara \& Nulsen 2007; Conroy
\etal\ 2015).  In addition, morphologies of the detected neutral gas
span a broad range, from regular disk- or ring-like structures to
irregular distributions of clumps and/or streams (e.g., Oosterloo
\etal\ 2007; Serra \etal\ 2012) with roughly 1/4 displaying
centralized disk or ring-like structures (Serra \etal\ 2012).  These
different morphologies indicate possibly different origins of the gas
in different galaxies, including left-over materials from previous
mergers and newly accreted gas from the CGM/IGM.  These gas-rich,
quiescent galaxies therefore provide an important laboratory for
studying the physical processes that regulate/terminate star formation
in the presence of cold gas in massive galaxies.

Incidentally, QSO absorption-line spectroscopy carried out in the
vicinities of red galaxies at intermediate redshift, $z\sim 0.5$,
continues to uncover extended cool halo gas to projected distances
beyond $d\approx 100$ kpc (e.g., Gauthier \etal\ 2009, 2010; Bowen \&
Chelouche 2011; Thom \etal\ 2012; Huang \etal\ 2016), demonstrating
that there may indeed be sufficient supplies of cool gas in these
massive quiescent halos.  Absorption spectroscopy along multiple
sightlines in the vicinities of lensing galaxies is particularly
interesting, because the small projected distances of these lensed QSO
sightlines from the lensing galaxies at $d=3-15$ kpc (or $d=1-2\,r_e$)
provide an unprecedented opportunity to probe and spatially resolve
the cool gas content both in the interstellar space and in the halos
of quiescent galaxies.

The fields are chosen because of the available high-quality optical
and near-infrared imaging data in the {\it Hubble Space Telescope}
({\it HST}) archive, which enable a detailed morphological study of
galaxies near the QSO sightlines.  In addition, we have targeted
relatively wide-separation gravitational lens systems with angular
separations $\theta\apg 1.5''$ and relatively bright lensed QSO images
with $g$-band magnitude of $g\apl 19.5$.  These selection criteria are
chosen to facilitate high-resolution echelle spectroscopy of the
lensed QSOs on the ground.  The wide separation ensures that the
lensed QSO images are not blended under typical sub-arcsecond seeing
conditions on the ground, allowing these lensed QSO images to serve as
independent probes of the foreground CGM at multiple locations.  The
brightness limit ensures a relatively high observing efficiency of QSO
echelle spectroscopy under limited observing resources.

As described in \S\ 3 below, all three lensing galaxies presented here
are old and massive with stellar masses in the range of
$\log\,M_*/M_\odot=10.9-11.4$. 
The available QSO echelle spectra cover a spectral range from $\approx
3300$ \AA\ to beyond 6000 \AA, which allow us to search for absorption
features due to Fe\,II, Mg\,II, Mg\,I, and Ca\,II transitions at the
redshift of each lensing galaxy.  These transitions are commonly seen
in the diffuse interstellar and circumgalactic gas of temperature
$T\sim 10^4$ K in the Milky Way and in distant galaxies (e.g., Savage
\& Sembach 1996; Rao et al.\ 2006).  The presence of such cool gas,
together with the expected presence of a hot X-ray emitting halo
around massive elliptical galaxies (e.g., O'Sullivan \etal\ 2001),
would indicate a multiphase nature of extended gas around these
lensing galaxies.  The observed velocity gradient across the lensing
galaxy, together with the relative abundance pattern between different
ions, also offers a unique opportunity to directly test the origin of
the observed cool gas in these massive, quiescent halos.

This paper is structured as follows. In Section 2, we describe
relevant imaging and spectroscopic observations and data reduction. In
Section 3, we describe the empirical properties that can be extracted
from available imaging and spectroscopic data.  Specifically, we
summarize the general observable quantities of the lensing galaxies in
\S\ 3.1 and absorbing gas properties in \S\ 3.2.  Observational
findings of individual fields are presented in Section 4, and analysis
of absorbing gas properties, including the ionization state, chemical
abundance pattern, and spatial coherence in gas kinematics, is
presented in section 5.  We discuss the implications for the origin of
chemically-enriched cool gas near massive, quiescent galaxies in
Section 6, and present a summary of our findings/conclusions in
Section 7.  We adopt a $\Lambda$CDM cosmology, $\Omega_{\rm M}=0.3$
and $\Omega_\Lambda = 0.7$, with a Hubble constant $H_0 = 70 \ {\rm
  km} \ {\rm s}^{-1}\ {\rm Mpc}^{-1}$ throughout the paper. All
magnitudes and colors reported here are in the AB system.

\section[]{Observations and Data Reduction}

The study presented in this paper focuses on three multiply-lensed QSO
fields, HE\,0047$-$1756, HE\,0435$-$1223, and HE\,1104$-$1805, with
the lensing galaxies identified at $z=0.4-0.7$.  Figure 1 shows the
lensing configurations of these fields.  The relatively wide
separation ($\theta_{\rm lens}\apg 1.5''$) enables the application of
these lensed QSO images as independent probes of the inner gaseous
halo around each of the lensing galaxies based on QSO absorption
spectra obtained on the ground.  Here we describe relevant imaging and
spectroscopic data of the lensing galaxies, as well as echelle
spectroscopy of the lensed QSOs.

\subsection{Imaging Observations}

Exquisite optical and near-infrared images of the lensed QSO fields
were retrieved from the {\it Hubble Space Telescope} (HST) data
archive.  
Details of the imaging observations are
summarized in Table 1. 
All imaging data were processed using the standard HST reduction
pipeline, and individual dithered exposures were drizzle-combined
using the AstroDrizzle package in each bandpass. Given slight
differences in WCS solutions between different bandpasses, the
co-added images were then registered to a common origin using point
sources in each field.  In Figure 1 we show false color composite
images of the three lens systems, where it can be seen that at the
redshift of each lens, absorption spectroscopy of the lensed QSOs
allows us to probe the gaseous halo on projected distance scales of
$d\sim 3-15$ kpc from the center of each lensing galaxy.

\begin{table}
\begin{center}
\caption{Journal of imaging observations with the HST}
\vspace{-0.5em}
\label{tab:HST}
\resizebox{3.4in}{!}{
\begin{tabular}{lccrrl}\hline
\multicolumn{1}{c}{Field}    	& Instrument 	& Filter & \multicolumn{1}{c}{Exptime}		& \multicolumn{1}{c}{PID} & \multicolumn{1}{c}{PI} \\	
 		&  		&       & \multicolumn{1}{c}{(s)}     &  &       \\\hline \hline
HE\,0047$-$1756 & ACS-WFC 	&F555W 	& 	670	&  9744 & C. Kochanek \\
		& ACS-WFC 	&F814W 	& 	670	&  9744 & C. Kochanek \\
  		& NICMOS 	&F160W 	& 	2620	&  9744 & C. Kochanek \\  \hline
HE\,0435$-$1223	& WFC3-UVIS 	&F275W 	& 	11360	&  11732  & C. Kochanek \\
  		& ACS-WFC 	&F555W 	& 	760	&  9744 & C. Kochanek \\
		& ACS-WFC 	&F814W 	& 	1440	&  9744 & C. Kochanek \\
  		& WFC3-IR	&F160W 	& 	9580	& 12889 & S. Suyu \\  \hline
HE\,1104$-$1805	& WFC3-UVIS 	&F275W 	& 	12620	&  11732  & C. Kochanek \\
  		& WFPC2-PC 	&F555W 	& 	9600	&  9138  & C. Impey \\
		& WFPC2-PC 	&F814W 	& 	8500	&  9138  & C. Impey \\
  		&WFC3-IR	&F160W 	& 	14380 & 12889 & S. Suyu\\  %\hline
\hline
\end{tabular}}
\end{center}
\end{table}

\subsection{Galaxy Spectra}

Optical spectra of two of the lensing galaxies, HE\,0047$-$1756 and
HE\,0435$-$1223, were published in Eigenbrod \etal\ (2006), and kindly
made available to us by F.\ Courbin.  A brief description of the
spectroscopic observations and spectra extraction is provided here.

The low-resolution spectra ($R\equiv\lambda/\Delta\,\lambda = 210$ at
5900 \AA) of the lenses were obtained using the FOcal Reducer and low
dispersion Spectrograph (FORS1) mounted on the European Southern
Observatory Very Large Telescope (ESO/VLT).  The observations were
carried out using a $1''$ slit under mean seeing conditions of ${\rm
FWHM}\approx 0.5"-0.6"$. Due to a significant amount of contaminating light in the galaxy
spectra from the lensed QSO images, Eigenbrod \etal\ (2006) employed a
spectral deconvolution algorithm to optimally extract the spectrum of
the lensing galaxy.  These authors applied the wavelength dependent
spatial profiles of known stellar (PSF) sources to deconvolve the
observed two-dimensional spectra of both the lensing galaxy and the
lensed QSO based on their known relative positions along the slit.
The extracted galaxy spectra show minimal residual presence of broad
emission features expected from the QSO, demonstrating the success of
the spectral deconvolution algorithm employed by Eigenbrod
\etal\ (2006).  Flux calibrations of the galaxy spectra were performed
using the same PSF stars adopted in the deconvolution routine.
Wavelengths were calibrated to air.

\begin{figure} 
\begin{center}
\includegraphics[width=85mm]{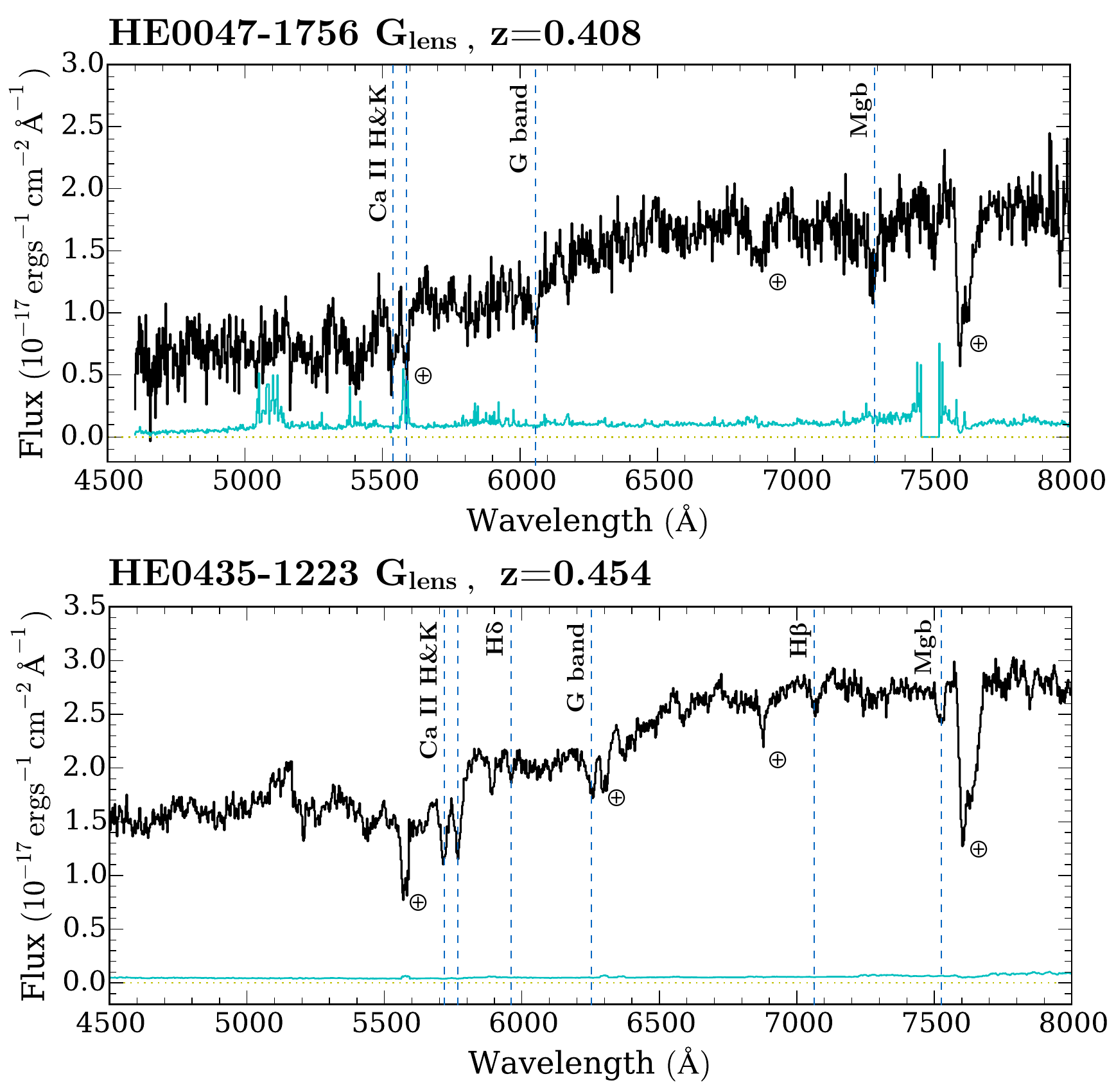}
\end{center}
\caption{Optical spectra of the lensing galaxies of HE\,0047$-$1756
  (top panel) and HE\,0435$-$1223 (bottom panel) at respective
  redshifts of $z=0.408\pm0.001$ and $z=0.454 \pm 0.001$ from Eigenbrod
  \etal\ (2006). The corresponding 1-$\sigma$ error spectrum of each
  galaxy is shown in cyan above the zero flux line (dotted line).  The
  galaxy spectra are dominated by prominent absorption lines such as
  Ca\,II H\&K, G-band, and Mg\,b as by dashed lines.  Night-sky
  absorption bands and emission line residuals are also marked by an
  earth symbol.}
\label{Figure 2}
\end{figure}

These optical spectra allow us to check the redshifts of
the galaxies based on vacuum wavelengths, providing the rest frame
necessary for studying halo gas kinematics using QSO absorption
spectroscopy, and to characterize the stellar population and star
formation history based on various line indices.  We present in Figure
2 the extracted spectra of the lensing galaxies for HE0047$-$1756 in
the top panel and for HE0435$-$1223 in the bottom panel, along with
their corresponding 1-$\sigma$ error spectra.  Both spectra show the
prominent absorption features such as Ca\,II H\&K doublet, G-band, and
Mg\,b, and a strong 4000-\AA\ spectral discontinuity that are
characteristic of a quiescent galaxy with no trace of on-going star
formation.  Using a cross-correlation analysis with model spectra
constructed from the eigen spectra of the Sloan Digital Sky Survey
(SDSS) spectroscopic galaxy sample (e.g.\ Chen \etal\ 2010), we
estimate the redshift of the lensing galaxy at $z=0.408\pm 0.001$ for
HE\,0047$-$1756, consistent with previous measurements of Eigenbrod
\etal\ (2006) and Ofek \etal\ (2006).  The lens redshift of
HE\,0435$-$1223 was estimated to be $z=0.454\pm 0.001$, consistent
with the measurements of Eigenbrod \etal\ (2006) and Morgan \etal\
(2005).  The redshift accuracy is clearly limited by the spectral
resolution and the broad absorption width of available spectral
features.  

We do not have the optical spectrum of the HE\,1104$-$1805 lens and
therefore adopt the published redshift of $z=0.729\pm 0.001$ from
Lidman \etal\ (2000) as the rest frame of the galaxy in the subsequent
analysis.

\begin{table}
\begin{center}
  \caption{Journal of QSO echelle spectroscopy}
\vspace{-0.5em}
\label{tab:MIKE,HIRES}
\resizebox{3.25in}{!}{
\begin{tabular}{lccrc}\hline
\multicolumn{1}{c}{QSO Image}    &$z_{\rm{em}}$ & Instrument 	& \multicolumn{1}{c}{Exptime} 	& Date \\	
 		&  	&         		& \multicolumn{1}{c}{(s)}         &        \\\hline \hline
HE\,0047$-$1756 A 	& 1.676	& MIKE	& 	7200	& 2013/11 \\
HE\,0047$-$1756 B	& 1.676	& MIKE 	& 	5700	& 2013/11 \\  \hline
HE\,0435$-$1223 A  	& 1.689	& MIKE	& 	5400	& 2013/11 \\  
HE\,0435$-$1223 B	& 1.689	& MIKE 	& 	5400	& 2013/11 \\
HE\,0435$-$1223 C  	& 1.689	& MIKE	& 	10800   & 2013/11 \\
HE\,0435$-$1223 D	& 1.689	& MIKE	& 	10800	& 2013/10 \\ \hline
HE\,1104$-$1805 A	& 2.305	& HIRES & 	19300	& 1997/02 \\ 
	& 	& UVES  & 	19000	& 2001/06 \\ 
HE\,1104$-$1805 B	& 2.305	& HIRES & 	51200	& 1997/02 \\
	& 	& UVES  & 	19000	& 2001/06 \\
\hline
\end{tabular}}
\end{center}
\end{table}

\subsection{QSO Echelle Spectroscopy}

High-resolution echelle spectra of the doubly lensed QSO
HE\,0047$-$1756 and the quadruply-lensed QSO HE\,0435$-$1223 were
obtained using the MIKE echelle spectrograph (Bernstein \etal\ 2003)
on the Magellan Clay Telescope.  Following Chen \etal\ (2014), echelle
spectroscopic observations were carried out in the fall semester of
2013 using a $1''$ slit and an aggressive binning of $2\times4$
(spatial\,$\times$\,spectral) during readout to increase observing
efficiency of these relatively faint QSOs.  The mean seeing conditions
over the observing period were $\approx 0.5''-0.7''$.  The observing
setup delivers a spectral resolution of ${\rm FWHM}\approx 12$ \kms\
over the spectral range from $\lambda=3200$ \AA\ to $\lambda=1\,\mu$m.
While the heavy binning does not allow us to fully resolve individual
resolution components, it is sufficient to resolve velocity structures
on scales as small as $\sim 10$ \kms.  A summary of MIKE observations
is presented in Table 2.

The MIKE spectra were reduced using a custom data reduction pipeline
described in Chen \etal\ (2014).  In summary, each raw spectral image
was bias-subtracted and corrected for pixel-to-pixel variations using
twilight flats obtained through a diffuser.  A ThAr comparison frame,
obtained immediately following every science exposure, was used to
create a two-dimensional wavelength map, which was corrected to vacuum
and heliocentric wavelengths.  For each echelle order, the QSO
spectrum was optimally extracted using a Gaussian profile that matched
the spatial profile of the QSO.  To prevent possible
contamination from neighboring QSO images (in the case of
HE\,0435$-$1223) that were rotated into the slit during individual
exposures, we made sure to mask the trace of neighboring objects
during the spectral extraction procedure.  Relative flux calibrations
were performed using a sensitivity function determined from
observations of a spectrophotometric standard star taken during the
same night as the QSO observations.  Individual flux-calibrated
echelle orders from different exposures were co-added and combined to
form a single continuous spectrum per QSO image.  Finally, the
combined spectrum was continuum normalized using a low-order
polynomial function that fits the spectral regions free of strong
absorption features.  In the spectral region around the systemic
redshifts of the lensing galaxies, the mean $S/N$ per resolution
element is found to be $\approx 20-50$ for the two QSO images of
HE\,0047$-$1756, and $S/N \approx 10-15$ per resolution element for
all four images of HE\,0435$-$1223.

\begin{center}
\begin{table*}
  \centering
    \caption{Summary of Galaxy Structural and Photometric Properties}
\resizebox{7.in}{!}{
    \begin{tabular}{@{}ccccrccccc@{}}
      \hline
      &  &{$r_e$} &     & \multicolumn{1}{c}{PA$^a$}  & $M_B$ &  & & &SFR(UV)$^b$  \\
      Galaxy&   \multicolumn{1}{c}{$n$} & \multicolumn{1}{c}{(kpc)} & \multicolumn{1}{c}{$b/a$}  & \multicolumn{1}{c}{($^{\circ}$)} &(mag) & $(g-r)_{\mathrm{rest}}$ & $\rm{log} \,M_*/M_\odot$ & $ L_B/L_*$&(${\rm M}_\odot\,/\,{\rm yr}$)   \\
      \hline
      \hline
      HE\,0047\,$\rm G_{ lens}$& $3.87\pm 0.55$ & $2.6\pm 0.6$ & $0.78\pm 0.02$ & $114\pm 3$ & $-20.7$ & $0.77$ & $10.9$&0.9& ...  \\
      HE\,0435\,$\rm G_{ lens}$& $3.51\pm 0.04$ & $4.4\pm 0.1$ & $0.82\pm 0.01$ & $176\pm 1$ & $-21.4$ & $0.74$ & $11.1$&1.7& $<0.01$ \\
      HE\,1104\,$\rm G_{ lens}$& $4.36\pm 0.17$ & $8.2\pm 0.2$ & $0.77\pm 0.01$ & $56\pm 2$ & $-22.4$ & $0.71$ & $11.4$  & 3.5& $<0.1$ \\
 
      \hline
     \multicolumn{9}{l}{$^a$Position angle of the major axis of the lens galaxy is measured North through East.}\\
      \multicolumn{9}{l}{$^b$Unobscured SFR based on the observed F275W flux limits and the SFR calibrator from Kennicutt \& Evans (2012).}\\
    \label{gal_table}
  \end{tabular}
}
\end{table*}
\end{center}

High-resolution echelle spectra of the doubly lensed QSO
HE\,1104$-$1805 were obtained using High Resolution Echelle
Spectrometer (HIRES; Vogt \etal\ 1994) on the Keck I Telescope.  The
observations were carried out in February 1997 (Rauch \etal\ 2001)
using a $0.86''$ slit that delivered a spectral resolution of ${\rm
  FWHM}\approx 6.6$ \kms\ over the spectral range from $\lambda=3600$
\AA\ to $\lambda=6100$ \AA.
Additional echelle observations of the QSOs were obtained using the
VLT UV-Visual Echelle Spectrograph (UVES) in June 2001 (Lopez \etal\
2007) and kindly provided to us by Dr.\ Sebasti\'an L\'opez.  The
final combined spectra cover a wavelength range from $\lambda=3050$
\AA\ to $\lambda=6850$ \AA\ with a spectral resolution of ${\rm
  FWHM}\approx 6.7$ \kms.  The mean $S/N$ for image $A$ is $S/N\approx
80$ per pixel and $S/N\approx 50$ per pixel for image $B$.
Comparisons of absorption line profiles between the continuum
normalized HIRES and UVES spectra show excellent agreement between the
two datasets.  The additional UVES spectra offer coverage for the
Ca\,II\,$\lambda\,3934$ absorption transition at the redshift of the
lensing galaxy.  A summary of the HIRES and UVES observations is also
included in Table 2.

\section[]{Lensing Galaxy and Absorbing Gas Properties}

A principal goal of our multi-sightline study is to examine
whether/how spatially resolved halo gas kinematics from
high-resolution QSO absorption spectroscopy correlate with the
observed properties of the lensing galaxy.  In this section, we
describe the galaxy properties and halo gas absorption properties that
were extracted from the data described in \S\ 2.

\subsection{Galaxy Properties}

The available imaging and spectroscopic data of the lensing galaxies
described in \S\S\ 2.1 and 2.2 allow us to examine in detail their
optical morphologies and stellar population, thereby constraining the
stellar structure and star formation history.

To analyze the morphologies of the lensing galaxies, we have developed
a custom computer program to obtain a best-fit two-dimensional surface
brightness profile of each of the lensing galaxies.  The results of
the surface brightness analysis include morphological parameters (such
as effective radius, ellipticity, etc.), and total flux.

The first step in our surface brightness analysis is a careful
subtraction of lensed QSO images, a necessity given the close angular
separations (typically $\approx 1''$) between the lensing galaxy and
the lensed QSO images.  The QSO image subtraction was performed using
a model point spread function (PSF) generated from the TinyTim
software (Krist 1995) for individual cameras.  We scale the input PSF
to match both the observed peak brightness and position of each lensed
QSO image, and subtract the best-fit point source model profile from
the QSO image.  Following the QSO image subtraction, we characterize
the two-dimensional surface brightness profile of each lensing galaxy,
$\mu$, as a function of galactocentric radius, $R$, following a
S\'ersic profile,
\begin{equation}
\mu\mathrm{(}R\mathrm{)}=\mu_0 \ {\rm exp}\left\{ -b_n \left[ \left( \frac{R}{r_e}\right)^\frac{1}{n} -1\right]\right\},
\end{equation}
where $\mu_0$ is the central surface brightness, $r_e$ is the
effective (also called half-light) radius, $n$ is the S\'ersic index,
and $b_n$ is a constant whose value is defined by $n$ ($b_n$ can be
precisely estimated by a polynomial in $n$, see Ciotti \& Bertin
1999).  The observed projected distance, $r_{ij}$, of each pixel $(i,j)$ in the
image is related to the deprojected galactocentric radius, $R_{ij}$,
according to,
\begin{equation}
R_{ij}=r_{ij}\,\sqrt{ 1 + \mathrm{sin^2}(\alpha_{ij}-{\rm PA})\,\mathrm{tan^2}(\cos^{-1}\,b/a)},
\end{equation}
where $\alpha_{ij}$ is the azimuthal angle of the pixel $(i,j)$ and PA is the
position angle of the major axis of the galaxy, both measured north
through east, and $b/a$ is the minor-to-major axis ratio.  The model
surface brightness profile is then convolved with a TinyTim PSF model
computed at the location of the lensing galaxy.  Finally, the
resulting PSF-convolved two-dimensional model surface brightness
profile is compared with the observation to obtain the best-fit
morphological parameters based on a $\chi^2$ analysis.  Error bars in
the best-fit parameters are estimated from the diagonal terms of the
covariance matrix returned from the $\chi^2$ analysis.

Of the images available in each field (Table 1), the F814W images
offer the best combination of fine spatial sampling and sufficient
sensitivity for tracing the dominant stellar population in the
rest-frame optical window.  While the two-dimensional surface
brightness profile analysis is performed for all bandpasses, we adopt
the best-fit morphological parameters as the fiducial model for
characterizing each lensing galaxy.  Cross examinations of the
best-fit morphological parameters in different bandpasses show that
the best-fit model based on the F814W image is consistent with those
obtained in other bandpasses for all three fields, except for the
F555W image of the HE\,1104$-$1805 lens.  The discrepancy is
understood as due to a lack of young stars contributing to the
rest-frame near-ultraviolet light that is recorded in the observed
F555W frame.  We present the best-fit morphological parameters from
the F814W images in columns (3) to (6) of Table 3.

\begin{center}
\begin{table}
 \centering
	\caption{Summary of Galaxy Spectroscopic Properties} \centering \label{table:spectroscopic} 
\resizebox{3.25in}{!}{
	\begin{tabular}{@{}clcccc@{}}  
	\hline  
		&  &  &  EW(H$\delta$)$^a$ & EW([O\,II])$^{b}$ & {${\rm SFR}{\rm [O\,II]}^{c}$} \\
		Galaxy & \multicolumn{1}{c}{$z$} & D4000 & (\AA) & (\AA) & (${\rm M}_\odot\,/\,{\rm yr}$)\\
		\hline \hline
		HE\,0047\,$\rm G_{ lens}$ & 0.408 & $1.4\pm 0.4$ & $2.8\pm 0.8$ & $<1.9$ & $<0.07$ \\
		HE\,0435\,$\rm G_{ lens}$ & 0.454 & $1.4\pm 0.2$ & $1.7\pm 0.2$ & $<0.4$ & $<0.06$ \\
		HE\,1104\,$\rm G_{ lens}$ &$0.729^d$ & ... & ... & ... & ... \\ %& ... &...&... \\
		\hline 
                \multicolumn{6}{l}{$^a$ Rest-frame absorption equivalent width} \\
                \multicolumn{6}{l}{$^b$ 2-$\sigma$ upper limit of [O\,II] emission.} \\
		\multicolumn{6}{l}{$^c$ Unobscured SFR based on the observed flux limits for [O\,II] emission} \\
                \multicolumn{6}{l}{\ \ line and the conversion from Kewley \etal\ (2004).} \\
                \multicolumn{6}{l}{$^d$ Lidman \etal\ (2000). }
	\end{tabular}
}	 
\end{table}
\end{center}

The total flux of each galaxy in each filter is determined by
integrating the best-fit S\'ersic profile within the half-light radius
$r_e$ and multiplying the result by a factor of two (yielding the
total flux, by definition of the S\'ersic profile)\footnote{It is
  possible that there exists an underlying faint disk that has been
  missed in the glare of the QSO images.  We show in the right most
  panel of Figures 3, 5, \& 7 that the 2-sigma limit on the surface
  brightness at $d > r_e$ of each lensing galaxy is typically fainter
  than $\mu \approx 23-23.5\,{\rm mag}/{\rm arcsec}^2$.  Taking a disk
  scale length of $r_s \approx 8$ kpc, which is typical of nearby
  elliptical/S0 galaxies (see e.g., de Jong \etal\ 2004), we estimate
  that the disk component, if present, would have added $< 35$\% of
  the total light.}.  Error in the total flux is estimated by adding
in quadrature the systematic error in the best-fit S\'ersic profile
due to uncertainties in the best-fit model parameters and statistical
errors due to photon counting.  Next, the rest-frame absolute
magnitudes and rest-frame $g-r$ colors, and total stellar mass $M_*$
of each galaxy are calculated using the IDL K-correct library
version\,4.2 (Blanton \& Roweis 2007), which performs
\textit{k}-correction using a library of templates generated using
Bruzual \& Charlot (2003) stellar population synthesis code together
with emission line models from Kewley \etal\ (2001).  For the lensing
galaxies in the HE\,0435$-$1223 and HE\,1104$-$1805 fields,
constraints on the unobscured star formation rate (SFR) are also
derived using the observed flux limit in the F275W bandpass and the
SFR calibrator from Kennicutt \& Evans (2012).  At the redshifts of
the lensing galaxies, the F275W bandpass corresponds to roughly
rest-frame $1500-1800$ \AA.  The results are also summarized in Table
3.

\begin{table*}
\centering
    \caption{Integrated Absorption Properties Along Multiple Sightlines near the Three Lensing Galaxies$^a$}
\resizebox{6.5in}{!}{
    \begin{tabular}{@{}ccrcccccccr@{}}
      \hline
      & & &  \multicolumn{2}{c}{Fe\,II\,$\lambda\,2600$} & \multicolumn{2}{c}{Mg\,II\,$\lambda\,2796$} & \multicolumn{2}{c}{Mg\,I\,$\lambda\,2852$} & \multicolumn{2}{c}{Ca\,II\,$\lambda\,3934$} \\
\cmidrule(lr){4-5} \cmidrule(lr){6-7} \cmidrule(lr){8-9} \cmidrule(lr){10-11}
      & \multicolumn{1}{c}{$\theta^b$} & \multicolumn{1}{c}{$d^c$}  & \multicolumn{1}{c}{$W_r$} & \multicolumn{1}{c}{$\delta\,v_{90}$} & \multicolumn{1}{c}{$W_r$} & \multicolumn{1}{c}{$\delta\,v_{90}$} & \multicolumn{1}{c}{$W_r$} & \multicolumn{1}{c}{$\delta\,v_{90}$} & \multicolumn{1}{c}{$W_r$} & \multicolumn{1}{c}{$\delta\,v_{90}$}  \\
      & ($''$)  & \multicolumn{1}{c}{(kpc/$r_e$)} & (\AA) & \multicolumn{1}{c}{(km/s)} & (\AA) & \multicolumn{1}{c}{(km/s)} & (\AA) & \multicolumn{1}{c}{(km/s)} & (\AA) & \multicolumn{1}{c}{(km/s)} \\
      \hline
      \hline
      \multicolumn{11}{c}{HE\,0047$-$1756 lens at $z=0.408$} \\
      \hline
      $A$ & 0.85 & 4.6/1.8 &  $2.29\pm0.04$ & $590$ & $4.46\pm0.02$ & $609$ & $0.80\pm0.02$ & $603$ & $0.32\pm 0.02$ & $584$ \\
     % \hline
      $B$ & 0.61 & 3.3/1.3   & $2.03\pm 0.10$ & $462$ & $3.69\pm0.04$ & $482$ & $0.72\pm0.04$ & $473$ & $0.12\pm 0.02$ & $74$ \\
      \hline 
      \multicolumn{11}{c}{HE\,0435$-$1223 lens at $z=0.454$} \\
      \hline
      $A$ & 1.30 & 7.5/1.7   & $<0.04$ & ... & $<0.03$ & ... & $<0.01$ & ... & $<0.03$ & ... \\
      %\hline
      $B$ & 1.17 & 6.7/1.5  & $<0.04$ & ... & $<0.03$ & ... & $<0.01$ & ... & $<0.04$ & ... \\
     % \aline
      $C$ & 1.30 & 7.5/1.7  & $<0.03$ & ... & $<0.03$ & ... & $<0.01$ & ... & $<0.03$ & ... \\
      %\aline
      $D$ & 1.07 & 6.2/1.4  & $<0.06$ & ... & $<0.04$ & ... & $<0.02$ & ... & $<0.03$ & ... \\
      \hline
      \multicolumn{11}{c}{HE\,1104$-$1805 lens at $z=0.729$} \\
      \hline
      $A$ & 1.11 & 8.1/1.0  & $0.34\pm0.01$ & $332$ & $0.64\pm0.01$ & $331$ & $0.09\pm0.01$ & $333$ & $0.06\pm0.01$ & 334 \\
       %\hline
      $B$ & 2.08 & 15.1/1.8   & $<0.01$ & ... & $<0.01$ & ... & $<0.01$ & ... & $<0.04$ & ... \\
      \hline %\hline
\multicolumn{11}{l}{$^a$For non-detections, we present a 2-$\sigma$ upper limit to the absorption equivalent width.  For HE\,1104$-$1805B, the limits are}\\ 
\multicolumn{11}{l}{\ \ \  estimated over the same velocity interval as in HE\,1104$-$1805A. For HE\,0435$-$1223 A,B,C,D, the limits are estimated } \\
\multicolumn{11}{l}{\ \ \  over twice the resolution element (10 km/s).} \\
\multicolumn{11}{l}{$^b$$\theta$: angular separation between the lens and lensed QSO image.} \\
\multicolumn{11}{l}{$^c$$d$: Projected distance in units of kpc or half-light radius $r_e$.} \\
%\multicolumn{12}{l}{$^d$$\phi$, relative azimuthal angle of lensed QSO image from the major axis of the lensing galaxy, measured north to east.} \\
\label{Absorption summary Table}
  \end{tabular}
}
\end{table*}

The available optical spectra of two of the lensing galaxies described
in \S\ 2.2 (Figure 2) exhibit prominent absorption features such as
Ca\,II H\&K doublet, G-band, and Mg\,b, and a strong 4000-\AA\
spectral discontinuity that are characteristic of a quiescent galaxy
with no on-going star formation.  We measure the D4000 index, which is
defined as the continuum flux ratio between two spectral regions,
$4000-4100$ \AA\ and $3850-3950$ \AA, bracketing the 4000-\AA\
discontinuity, ${\rm D4000}\equiv f_{4000-4100}/f_{3850-3950}$, as
well as the H$\delta$ absorption and [O\,II] emission equivalent width
following the spectral index definitions of Balogh \etal\ (1999).  The
results are summarized in Table 4.  Based on the observed equivalent
width upper limits for [O\,II] emission, we apply the SFR conversion
of Kewley \etal\ (2004) and place a 2-$\sigma$ upper limit for the
unobscured SFR at $\approx 0.07\,M_\odot/{\rm yr}$.  In addition, the
relatively large D4000 indices indicates that the two lensing galaxies
have a minimum stellar age of $\apg 1 $ Gyr (e.g., Kauffmann \etal\
2003).  We note that the presence of residual QSO flux blueward of the
4000-\AA\ spectral discontinuity would result in underestimated D4000
indices, implying an even older stellar population for these galaxies.

\subsection{Absorbing Gas Properties}

The echelle spectroscopy described in \S\ 2.3 produces high-resolution
absorption spectra of gas along multiple sightlines around three
lensing galaxies.  These spectra enable a detailed study of how halo
gas properties around massive quiescent galaxies vary with different
physical locations.  At the redshifts of the three lensing galaxies,
the echelle spectra provide a wavelength coverage for observing
prominent absorption transitions Fe\,II\,$\lambda\,2600$,
Mg\,II\,$\lambda\lambda\,2796, 2803$, Mg\,I\,$\lambda\,2852$, and
Ca\,II\,$\lambda\lambda\,3934$.

To characterize halo gas around the lensing galaxies, we perform two
sets of measurements following what is described in Chen
\etal\ (2014).  First, we measure the total, integrated absorption
equivalent widths along individual sightlines.  Measurements of total
integrated absorption equivalent widths along individual sightlines
offer a baseline comparison between absorbers residing in massive
galaxy halos and the general absorber population from random sightline
surveys (e.g.\ Rao \etal\ 2006; Zhu \& M\'enard 2013; Seyffert
\etal\ 2013).  Next, we perform a Voigt profile analysis on a
component-by-component basis to constrain the gas column densities and
Doppler parameters of individual components.  Measurements of
individual absorption components allow us to examine detailed
kinematic structures and investigate possible variations in relative
ionic abundances both along and across individual sightlines.

To carry out the Voigt profile analysis, we developed a custom
software to analyze both heavily binned MIKE spectra and unbinned UVES
and HIRES data.  For each observed absorption line system, we first
generate a theoretical absorption line profile based on the minimum
number of discrete components, $n_c$, that is needed to explain the
observed Mg\,II absorption kinematics.  We focus on the Mg\,II
absorption doublet for setting the number of necessary components,
because among all observable features (Fe\,II, Mg\,II, Mg\,I, and
Ca\,II absorption) the lines of the doublet are expected to be the
strongest features in diffuse gas.  For each absorbing component, the
Voigt profile is uniquely defined by three free parameters: the
velocity offset of the line center relative to the systemic redshift
of the lensing galaxy ($v_c$), the absorption column density of ion X
($\log\,N_c[{\rm X}]$), and the Doppler parameter ($b_c$).  Next, the
theoretical absorption profile is convolved with a Gaussian line
spread function with the ${\rm FWHM}$ set by the appropriate
instrumental resolution, which is ${\rm FWHM}\approx 12$ \kms\ for
MIKE and ${\rm FWHM}\approx 6.7$ \kms\ for UVES and HIRES.  Following
this step, the convolved model Voigt profile is binned according to
the adopted spectral binning of the data.  Finally, the resulting
binned model spectrum is compared with the observed absorption
spectrum, and the best-fit parameters are determined based on a
$\chi^2$ analysis.  We perform the Voigt profile fitting procedure to
all available transitions of a given absorption system.  The velocity
offsets of individual components are fixed across all transitions,
whereas $\log\,N_c$ and $b_c$ are allowed to vary freely for different
transitions.  The results of the Voigt profile analysis also allow us
to determine the total velocity widths that enclose 90\% of total
optical depth, which enable a direct comparison of the mean velocity
field across different sightlines.  Finally, we note that because both
HIRES and UVES spectra are available for HE\,1104$-$1805, the best-fit
model parameters were found by a simultaneous $\chi^2$ fit to the
HIRES and UVES spectra.  The two spectra were combined for display
purposes only (Figure 8).

A summary of the integrated absorption properties along individual
sightlines is presented in Table 5.  For each lensed QSO sightline, we
present its angular separation $\theta$ and the corresponding
projected distance $d$ to the lensing galaxies, the relative azimuthal
angle $\phi$ of the lensed QSO image from the major axis of the lens,
the rest-frame absorption equivalent width ($W_r$), and velocity width
enclosing 90\% of the total optical depth ($\delta\,v_{90}$) of each
absorption transition. Uncertainties in $\delta\,v_{90}$ are of order
the size of the spectral pixel, which is 10 \kms\ for HE\,0047$-$1756
and HE\,0435$-$1223, and 2.6 \kms\ for HE\,1104$-$1805.

\begin{figure*} 
\begin{center}
\includegraphics[width=160mm]{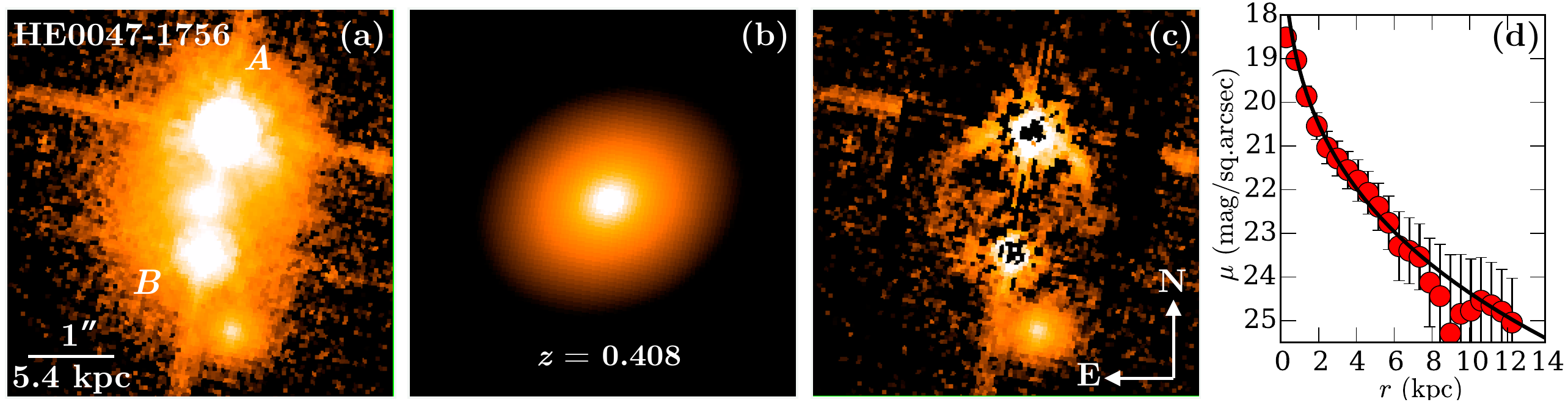}
\end{center}
\caption{(a) HST/ACS F814W image of the field around HE\,0047$-$1756.
  The lensing galaxy at $z=0.408$ is at the center of the panel
  surrounded by the doubly-lensed QSO at $z_{\rm QSO}=1.67$ (Wisotzki
  \etal\ 2004).  We note that the redshift of the fainter source at
  $1.5''$ south of the lens is not known.  Panel (b) displays the
  best-fit S\'ersic model surface brightness profile, which is
  characterized by a S\'ersic index of $n=3.9\pm 0.6$, half-light
  radius $r_e=2.6\pm 0.6$ kpc, and an axis ratio of $b/a=0.78\pm
  0.02$. The best-fit morphological properties are consistent with
  those expected from an elliptical galaxy.  Panel (c) displays the
  residual image after subtracting the QSO PSFs and the best-fit
  S\'ersic model. Panel (d) displays an azimuthally-averaged surface
  brightness profile of the lensing galaxy in the F814W bandpass.  Each
  data point is calculated from an elliptical annulus of 0.1" (2
  pixels) in width, and the associated error bar represents photon
  counting noise, which is driven primarily by the contaminating QSO
  light at large radii.  The best-fit S\'ersic from the 2D analysis is
  shown as a black line.  We note that the deviant points at $8-10$ kpc
  are due to large residuals from subtracting the QSO light.}
\label{Figure 3}
\end{figure*}

The results of the Voigt profile analysis for individual fields are
summarized in Tables 6, 7, and 8, and the best-fit model absorption
profiles are presented in Figures 4, 6, and 8 below for comparison
with observations.  Uncertainties listed in the tables represent only
formal errors from the $\chi^2$ analysis, not including systematic
uncertainties due to continuum fitting errors.  For non-detected
Fe\,II, Mg\,I, and Ca\,II absorbing components, we first measure the
2-$\sigma$ equivalent width limits of the strongest transitions using
the error spectrum over a spectral window that is twice the FWHM of
the corresponding Mg\,II component.  When Mg\,II absorption features
are absent (namely for HE\,0435$-$1223 and for HE\,1104$-$1805$B$),
the upper limits are evaluated over a spectral window that is twice
the spectral resolution element.  Then we calculate the corresponding
2-$\sigma$ upper limits to the component column densities, assuming
that the gas is optically thin.

Because our MIKE spectra have a pixel resolution comparable to the
instrument spectral resolution, it is important to understand and
quantify possible systematic uncertainties in the Voigt profile
analysis that may result from an undersampled line spread function.
For this reason, we perform a series of Monte Carlo simulations to
assess the accuracy of the best-fit Voigt profile parameters.
Specifically, we first generate a set of synthetic Mg\,II and Fe\,II
absorption lines with varying column densities and Doppler parameters,
convolve the synthetic absorbers with the instrumental line spread
function, apply a pixel binning of $10$ \kms, and add noise to the
synthetic spectra based on the 1-$\sigma$ error spectrum associated
with each lensed QSO spectrum.  Next, we perform the Voigt profile
analysis using the resulting synthetic spectrum to determine the
best-fit column density and Doppler parameter of each component.
Finally, we repeat the process 1000 times to record the distribution
of best-fit model parameters relative to the input values.

The Monte Carlo simulations demonstrate that despite an undersampled
line spread function, the input column density and Doppler parameter
are well-recovered for relatively broad components ($\apg 7$ \kms).
For these broad components, the 95\% confidence interval is less than
5\% of the best-fit column density and $b$ values.  For narrower
components ($b<7$ \kms), we found that while the input values are
well-recovered for weak transitions ($\log\,N\apl 13.0$), the
intrinsic line profiles begin to saturate when the column density
exceeds $\log\,N\approx 13.0$.  The simulations show that for these
narrow and saturated components the best-fit column density and $b$
value become degenerate and the uncertainties in column density can be
as high as 0.1 dex for absorbers of $\log\,N\apl 13.5$ and up to 0.3
dex for strong absorbers of $\log\,N>14$.  For Mg\,II components, we
note that all narrow component are weak and saturation is therefore
not a significant issue.  For Fe\,II components, including weaker
transitions, such as Fe\,II\,$\lambda\,2586$, allows us to recover the
underlying $N({\rm Fe\,II})$ to better than 0.1 dex accuracy for
components as strong as $\log\,N({\rm Fe\,II})\approx14$.  Based on
the results of the simulations, we conclude that the column density
measurements are robust for relatively isolated components.

Complications arise when two saturated components are blended
together.  This is the case for components 1 \& 2 in sightline $B$ of
HE\,0047$-$1756 (see Figure 4 below).  A simultaneous Voigt profile
analysis of the two components yields two local $\chi^2$ minima at
$(\log\,N_c,b_c) = (16.5, 6.9)$ and (13.7, 18.0) for component 1, and
$(\log\,N_c,b_c) = (16.8, 14.9)$ and (14.9, 23.2), for component 2.
For these two components, we apply a prior based on the known $N({\rm
  Mg\,I})$ and the Cloudy photo-ionization calculation (see \S\ 5.1,
and adopt the second local minimum as the best-fit values.  This is
justified by the implied $N({\rm Mg\,I})/N({\rm Mg\,II})$ ratio which,
at the first local minimum, would place the component in an
unrealistically low gas density regime, $n_{\rm H}<10^{-4}\,{\rm
  cm}^{-3}$ (see Figure 9 below), leading to an unphysically large
cloud size that exceeds 300 kpc for optically thin gas with solar
metallicity.

\section[]{Description of Individual Lensing Galaxies}

With the separate measurements of galaxy and absorption-line
properties presented in \S\ 3, here we proceed with a joint analysis
of the stellar population and halo gas properties of each lensing
galaxy.

\subsection{The HE\,0047$-$1756 Lens at $z=0.408$}

The lensing galaxy of HE\,0047$-$1756 at $z=0.408$ was
spectroscopically identified by Ofek \etal\ (2006) and confirmed by
Eigenbrod \etal\ (2006).  Our two-dimensional surface brightness
profile analysis has yielded a best-fit S\'ersic index of $n=3.9\pm
0.6$, a half-light radius of $r_e=2.6$ kpc, and an axis ratio of
$b/a=0.78\pm 0.02$ (Figure 3).  The best-fit morphological parameters
characterize the lens as an elliptical galaxy, which is consistent
with the relatively old age ($>1$ Gyr) and a lack of on-going star
formation (${\rm SFR}<0.07\,{\rm M}_\odot\,{\rm yr}^{-1}$) inferred
from the spectral indices presented in Table 4.  Integrating the
best-fit S\'ersic profile, we estimate the total apparent magnitudes
in the F555W, F814W, and F160W bandpasses, and find
$AB\rm{(F555W)}=21.64\pm 0.07$, $AB\rm{(F814W)}= 19.71\pm0.05$, and
$AB\rm{(F160W)}=18.64\pm 0.06$.  Following the procedures described in
\S\ 3.1, the observed apparent magnitudes translate to a rest-frame
$B$-band absolute magnitude of $M_B=-20.7$, which is roughly $0.9\,L_*$ at
$z=0.4$ according to Faber \etal\ (2007).  In addition, we find a
rest-frame optical color of $g-r= 0.77$ and a total stellar mass of
$\log\,M_*/M_\odot=10.9$ for the lensing galaxy.  Adopting the
stellar-to-halo-mass relations of Behroozi \etal\ (2013) and Kravtsov
\etal\ (2014), we further infer a total dark matter halo mass of
$\log\,M_h/M_\odot=12.4-12.7$.  In summary, we find that the lensing
galaxy of HE\,0047$-$1756 is a quiescent $L_*$ galaxy with structural
and photometric properties typical of intermediate-redshift early-type
galaxies (e.g., Rutkowski \etal\ 2012).

We note the presence of an extended source at $\approx 1.6''$
southwest of the lensing galaxy.  This object appears to be bluer than
the lensing galaxy (see left panel of Figure 1).  We have also
performed a two-dimensional surface brightness profile analysis for
this object and found a best-fit S\'ersic index of $n=2.05\pm0.08$,
consistent with the more extended morphology displayed in the HST
images.  No redshift measurement is available for this object.
However, Chantry \etal\ (2010) noted that this galaxy is likely a
major contributor to the shear term in the lens model that is needed
to reproduce the image configuration of the lensed QSOs.  Assuming
this object is at the same redshift as the lensing galaxy, Chantry
\etal\ (2010) calculated a velocity dispersion of $\sigma=88$
\kms\ for this galaxy, which is comparable to the characteristic
velocities of large satellite galaxies like the Large Magellanic Cloud
(e.g., Alves \& Nelson 2000).  

At $z=0.408$, the projected distance between the lens and QSO image $A$
is $d_\mathrm{A}=4.6$ kpc or $1.8\,r_e$, and the projected distance
between the lens and QSO image $B$ is $d_\mathrm{B}=3.3$ kpc or
$1.3\,r_e$.  The two sightlines probe both the gaseous halo at small
projected distances and the interstellar medium of an elliptical
galaxy where the gas is expected to be hot.  Adopting the correlation
between X-ray luminosity and $B$-band luminosity $L_B$ of local
elliptical galaxies from O'Sullivan \etal\ (2001), we infer an X-ray
luminosity of $L_X\approx 10^{41}\,{\rm erg}\,{\rm s}^{-1}$ for the
lens of HE\,0047$-$1756.  The expected luminous X-ray flux indicates
that, similar to nearby elliptical galaxies, the lensing galaxy is
likely surrounded by a hot halo.

\begin{figure} 
\begin{center}
\includegraphics[width=84mm]{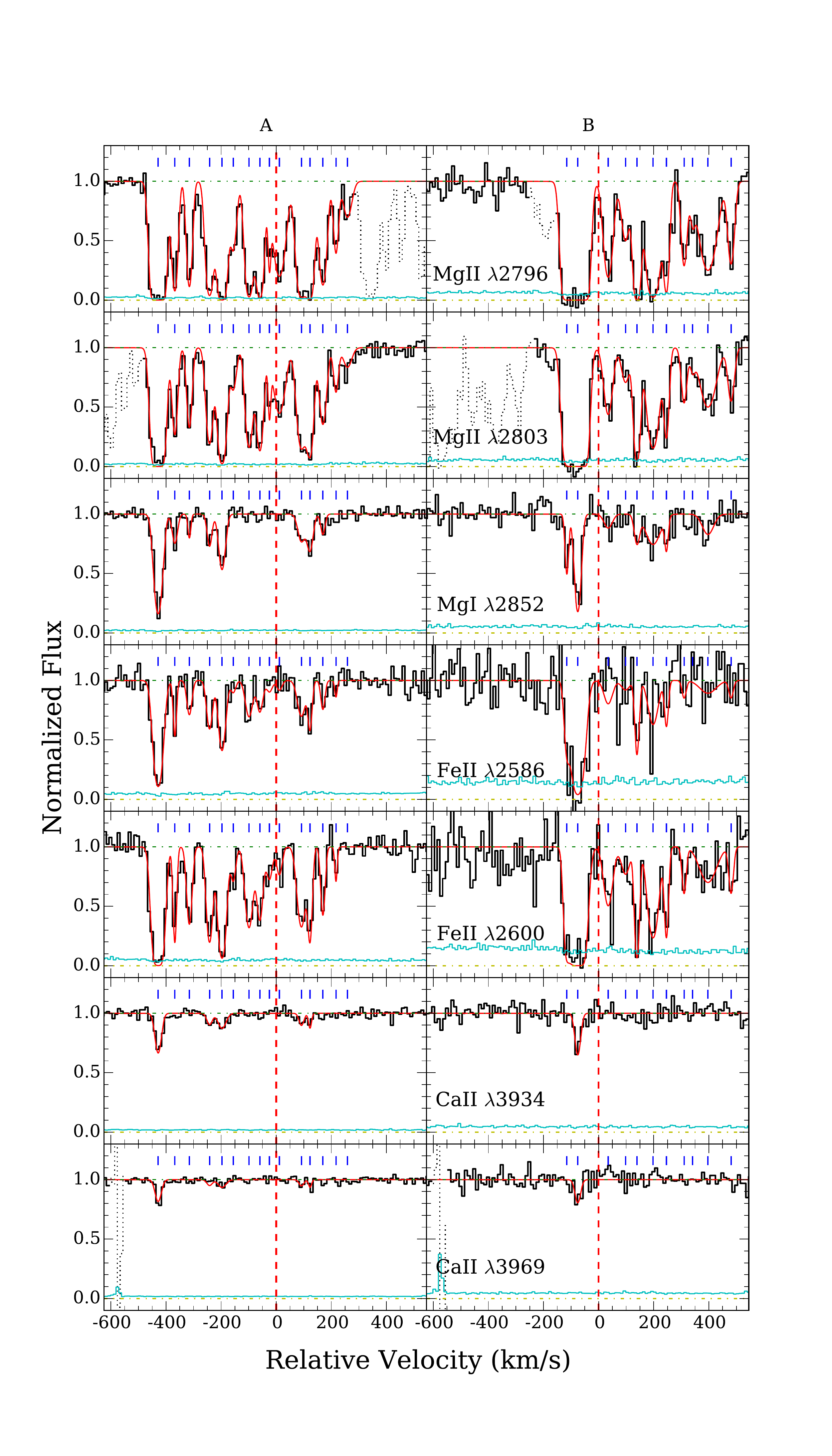}
\end{center}
\caption{Continuum normalized absorption profiles of different
  transitions along the lensed QSO sightlines HE\,0047$-$1756 $A$
  (left) at $d=4.6$ kpc (or $1.8\,r_e$) and $B$ (right) at $d=3.3$ kpc
  (or $1.3\,r_e$) from the HE\,0047$-$1756 lens.  Zero velocity
  corresponds to the systemic redshift of the lensing galaxy at
  $z=0.408$.  The 1-$\sigma$ error spectrum is included in thin, cyan
  curve above the zero flux level.  The blue tickmarks at the top of
  each panel indicate the location of individual components included
  in the Voigt profile analysis (see \S\ 3.2), and the best-fit Voigt
  profile models are included in red.  Contaminating features have
  been dotted out for clarity.  Despite showing no sign of recent star
  formation, the inner halo of the lensing galaxy harbors a
  significant amount of cool gas with complex kinematic profiles that
  span $\approx 500-600$ \kms\ in line-of-sight velocity and exhibit a
  large velocity shear of $\Delta\,v\approx 350$ \kms\ across the two
  sightlines separated by merely $\approx 8$ kpc. }
\label{Figure 4}
\end{figure}

\begin{table*}
\scriptsize
\centering
    \caption{Absorption properties around the HE\,0047$-$1756 Lens.}
    \begin{tabular}{ccrrcrcrcrcr}
      \hline
      & & & \multicolumn{1}{c}{$v_c$} & \multicolumn{2}{c}{Fe\,II} & \multicolumn{2}{c}{Mg\,II} & \multicolumn{2}{c}{Mg\,I} & \multicolumn{2}{c}{Ca\,II} \\
\cmidrule(lr){5-6} \cmidrule(lr){7-8} \cmidrule(lr){9-10} \cmidrule(lr){11-12}
      Sightline  & \multicolumn{1}{c}{$n_c$} & \multicolumn{1}{c}{component} & \multicolumn{1}{c}{(km/s)} &  $\log\,N_c$ & \multicolumn{1}{c}{$b_c$ (km/s)} &  $\log\,N_c$ & \multicolumn{1}{c}{$b_c$ (km/s)} &  $\log\,N_c$ & \multicolumn{1}{c}{$b_c$ (km/s)} &  $\log\,N_c$ & \multicolumn{1}{c}{$b_c$ (km/s)} \\
      \hline
      \hline
      $A$   & 15 &  1 &  $-428.4$ & $14.23\pm 0.03$ & $19.1\pm 0.7$ & $14.34\pm 0.16$ & $19.2\pm 1.4$ & $12.63\pm 0.02$ & $17.1\pm 0.8$ & $12.23\pm 0.03$ & $14.3\pm 1.4$ \\
          &    &  2 &  $-367.9$ & $13.32\pm 0.08$ &  $5.5\pm 1.1$ & $13.12\pm 0.03$ & $11.0\pm 0.9$ & $11.68\pm 0.21$ & $11.7\pm 0.2$ & $<11.0$ &   ... \\
          &    &  3 &  $-315.2$ & $13.21\pm 0.05$ & $11.0\pm 1.6$ & $13.04\pm 0.03$ & $10.8\pm 0.8$ & $11.35\pm 0.12$ &  $4.4\pm 5.4$ & $<11.1$ &   ... \\
          &    &  4 &  $-241.9$ & $13.45\pm 0.04$ & $12.8\pm 1.3$ & $13.31\pm 0.03$ & $14.5\pm 1.3$ & $11.69\pm 0.08$ & $10.4\pm 3.4$ & $11.57\pm 0.12$ & $12.3\pm 5.5$ \\
          &    &  5 &  $-196.6$ & $13.75\pm 0.03$ & $15.5\pm 1.2$ & $13.70\pm 0.06$ & $15.5\pm 1.7$ & $12.11\pm 0.04$ & $14.7\pm 1.8$ & $11.86\pm 0.08$ & $18.9\pm 4.8$ \\
          &    &  6 &  $-155.9$ & $12.70\pm 0.13$ & $10.7\pm 5.0$ & $12.71\pm 0.08$ & $13.9\pm 3.4$ &     $<10.9$     &     ...       & $<11.0$ &   ... \\
          &    &  7 &   $-98.6$ & $13.39\pm 0.04$ & $16.7\pm 2.0$ & $13.37\pm 0.03$ & $15.2\pm 1.3$ &     $<10.9$     &     ...       & $<11.1$ &   ...  \\
          &    &  8 &   $-59.3$ & $13.22\pm 0.06$ & $12.9\pm 2.7$ & $13.39\pm 0.03$ & $14.1\pm 1.3$ &     $<10.9$     &     ...       & $<11.1$ &   ...  \\
          &    &  9 &   $-24.7$ & $12.75\pm 0.16$ & $13.2\pm 7.2$ & $12.78\pm 0.27$ &  $3.2\pm 1.4$ &     $<10.7$     &     ...       & $<10.8$ &   ...  \\
          &    & 10 &   $+11.4$ & $12.64\pm 0.15$ & $12.0\pm 6.8$ & $13.22\pm 0.03$ & $25.7\pm 2.0$ &     $<11.0$     &     ...       & $<11.2$ &   ... \\
          &    & 11 &   $+91.5$ & $13.42\pm 0.04$ & $18.4\pm 2.3$ & $13.47\pm 0.09$ & $19.0\pm 2.8$ & $11.77\pm 0.10$ & $16.4\pm 4.9$ & $11.55\pm 0.12$ & $10.6\pm 4.9$ \\
          &    & 12 &  $+122.6$ & $13.31\pm 0.06$ &  $8.2\pm 1.6$ & $13.46\pm 0.09$ & $13.6\pm 2.3$ & $11.85\pm 0.07$ & $13.7\pm 2.8$ & $11.45\pm 0.12$ &  $4.8\pm 6.7$ \\
          &    & 13 &  $+169.4$ & $13.04\pm 0.06$ &  $8.3\pm 2.0$ & $13.13\pm 0.03$ & $15.4\pm 1.3$ & $11.42\pm 0.16$ &  $9.0\pm 7.0$ & $<11.1$ & ... \\
          &    & 14 &  $+216.9$ & $12.62\pm 0.10$ &  $5.3\pm 2.2$ & $12.64\pm 0.05$ & $10.2\pm 2.1$ &     $<10.8$     &     ...       & $<11.1$ & ...  \\
          &    & 15 &  $+258.8$ &     $<12.2$     &      ...      & $12.46\pm 0.09$ & $19.5\pm 5.1$ &     $<10.9$     &     ...       & $<11.1$ & ... \\
      \hline
      $B$   & 11 &  1 &  $-115.3$ & $13.60\pm 0.16$  & $10.9\pm 3.5$ &    $13.74^{+0.10}_{-0.08}$  &  $18.0\pm 1.6$   & $11.92\pm 0.11$ &  $6.3\pm 3.2$ & $<11.5$ & ... \\
          &    &  2 &   $-75.6$ & $14.49\pm 0.12$  & $24.5\pm 2.6$ &    $14.88^{+0.29}_{-0.23}$  & $23.2\pm 2.0$   & $12.51\pm 0.06$ & $13.1\pm 1.6$ & $12.21\pm 0.05$ & $12.6\pm 2.4$ \\
          &    &  3 &   $+34.7$ & $13.24\pm 0.16$  & $20.1\pm 6.2$ & $13.11\pm 0.04$ & $19.0\pm 2.2$ & $11.52\pm 0.27$ & $19.7\pm 8.9$ & $<11.5$ & ... \\
          &    &  4 &   $+97.7$ & $12.82\pm 0.20$  & $20.0\pm 4.5$ & $12.72\pm 0.09$ & $18.4\pm 5.6$ &    $<11.3$      &    ...        & $<11.5$ & ... \\
          &    &  5 &  $+139.1$ & $13.59\pm 0.14$  &  $8.1\pm 4.2$ & $13.55\pm 0.10$ & $12.0\pm 1.5$ & $11.68\pm 0.20$ & $11.9\pm 4.6$ & $<11.4$ & ... \\
          &    &  6 &  $+197.6$ & $13.62\pm 0.07$  & $23.0\pm 4.2$ & $13.57\pm 0.04$ & $24.8\pm 2.7$ & $12.12\pm 0.12$ & $34.3\pm 6.8$ & $<11.6$ & ... \\
          &    &  7 &  $+246.1$ & $13.25\pm 0.17$  &  $7.3\pm 3.8$ & $13.12\pm 0.07$ & $10.2\pm 1.8$ & $11.59\pm 0.21$ &  $6.4\pm 7.6$ & $<11.4$ & ... \\
          &    &  8 &  $+310.2$ & $12.80\pm 0.18$  &  $8.4\pm 6.2$ & $12.82\pm 0.06$ & $12.1\pm 2.6$ &    $<11.2$      &    ...        & $<11.5$ & ... \\
          &    &  9 &  $+340.9$ &     $<12.6$      &      ...      & $12.22\pm 0.20$ & $10.0\pm 6.0$ &    $<11.1$      &    ...        & $<11.4$ & ... \\
          &    & 10 &  $+396.8$ & $13.24 \pm 0.12$ & $40.0\pm 6.8$ & $13.30\pm 0.03$ & $36.4\pm 3.7$ & $11.83\pm 0.15$ & $27.4\pm 6.4$ & $<11.7$ & ... \\
          &    & 11 &  $+480.7$ & $12.83 \pm 0.18$ &  $9.5\pm 7.5$ & $12.85\pm 0.05$ & $13.9\pm 2.4$ &    $<11.2$      &    ...        & $<11.5$ & ... \\
      \hline %\hline
    \label{HE0047 Table}
  \end{tabular}
%\end{minipage}
\end{table*}
%\end{center}

At the same time, the absorption spectra show that not only abundant
cool gas is present along both $A$ and $B$ sightlines in the inner
halo of the galaxy, but the velocity spread is also very large.  The
total rest-frame equivalent width is found to be $W_r\rm{(2796)}=4.46
\pm 0.02$ \AA\ along sightline A at $d=4.6$ kpc ($1.8\,r_e$) north of
the lens and $W_r\rm{(2796)}=3.69 \pm0.04$ \AA\ along sightline B at
$d=3.3$ ($1.3\,r_e$) kpc south of the lens.  These ultra-strong
absorbers are often attributed to starburst driven outflows (e.g.,
Nestor \etal\ 2011).  In the case of the lensing galaxy, however, the
lack of on-going star formation together with a dominant old stellar
population makes a starburst driven outflow origin an unlikely
scenario.  On the other hand, Rao \etal\ (2006) showed that strong
Mg\,II absorbers of $W_r(2796)>0.6$ \AA\ at $z< 1.65$ have a mean H\,I
column density of $\langle\,N({\rm H\,I})\,\rangle=(3.5\pm 0.7) \times
10^{20}\,\cmjj$ which, together with the relative line ratios between
Mg\,II, Fe\,II, and Mg\,I, indicate a high probability ($>40$\%) that
the strong Mg\,II absorbers found in the inner regions of the lensing
galaxy are damped \lya\ absorbers (DLAs), in which the gas is expected
to be mostly neutral (e.g., Wolfe \etal\ 2005).

It is clear from the echelle absorption spectra that the large
equivalent widths are driven by complex multi-component structures
that span $\approx 500-600$ \kms\ in line-of-sight velocity (Table 5
and Figure 4).  The observed velocity spread along an individual
sightline exceeds the maximum circular velocity, $v_{\rm max}\approx
240$ \kms, expected for halos of $\sim\, 2.5\times10^{12}\,M_\odot$.
In addition, the absorption profiles exhibit an edge-leading signature
commonly seen in rotating disks with the highest column density gas
moving (blueshifted) at the highest velocity toward the observer.
Based on a grid of photo-ionization models discussed in \S\ 5.1 below,
we find that component 1 along sightline $A$ and component 2 along
sightline $B$ are indeed likely strong Lyman limit absorbers (or
possibly DLAs) of $\log\,N({\rm H\,I})>19$ which have a significant
neutral fraction.  Other weaker components remain in the optically
thin regime with $\log\,N({\rm H\,I})\apl 17$.  The lower $N({\rm
  H\,I})$ is also reflected in the observed declining $N({\rm
  Mg\,II})$ (by nearly 2 dex) with increasing receding velocity along
the line of sight.  Such edge-leading kinematic signatures are present
along both sightlines.

At $d=3-5$ kpc, the QSO sightlines probe both halo gas at $r\sim 100$
kpc (in front of and behind the lens) and the ISM at $r\sim 5-10$ kpc.
It is possible that a coherent structure in the ISM contributes
predominantly to the absorption profiles, which drives the strong
resemblance in the edge-leading signature, and that independent gas
clumps in the halo contribute to the absorption separately along
different sightlines, which adds noise to the profiles.  To examine
possible spatial coherence between the gas revealed along the two
sightlines, we perform a cross-correlation analysis of the absorption
profiles observed along $A$ and $B$ sightlines and found a clear
maximum at velocity offset of $\Delta\,v\approx 350$ \kms\ between the
two sightlines.  Applying an velocity offset of $\Delta\,v=-350$ \kms,
we find good match between components 1, 2, 5, 7, and 8 along
sightline $B$ and components 1, 5, 7, and 9 along sightline $A$, both
in their relative absorption strengths and in velocity offsets.  The
observed kinematic profiles suggest strong coherence on scales of
$\sim 8$ kpc in the gas motion across the inner halo (a more complete
discussion follows in \S\ 5.3).  However, the matched components span
a light-of-sight velocity range of $\approx 420$ \kms, still exceeding
the maximum circular velocity, $v_{\rm max}\approx 240$ \kms, expected
for halos of $\sim\, 2.5\times10^{12}\,M_\odot$.

\begin{figure*} 
\begin{center}
\includegraphics[width=160mm]{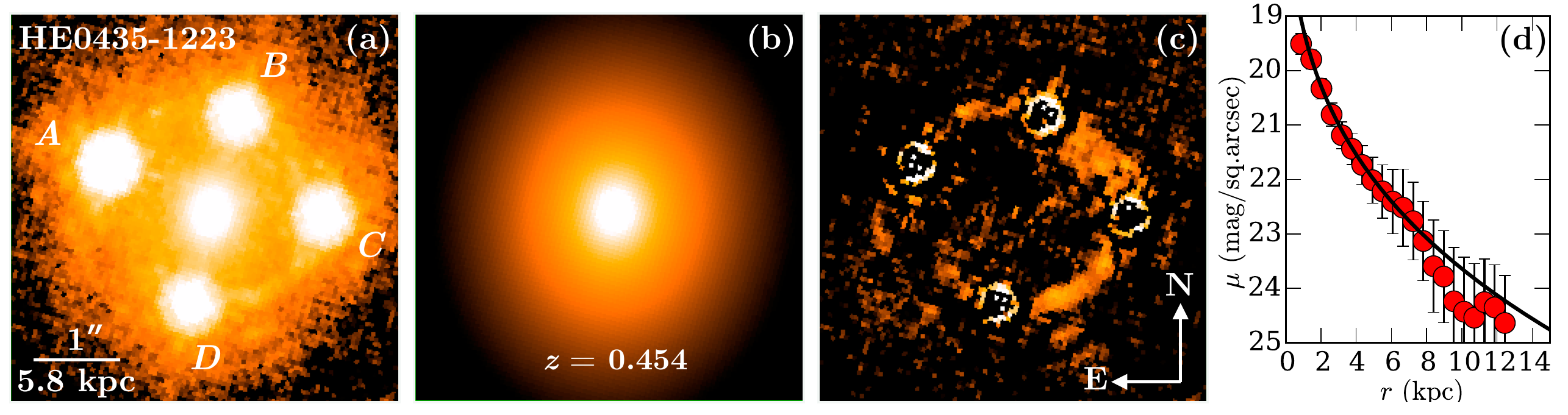}
\end{center}
\caption{(a) HST/ACS F814W image of the field around HE\,0435$-$1223.
  The lensing galaxy at $z=0.454$ is at the center of the panel
  surrounded by the quadruply-lensed QSO at $z_{\rm QSO}=1.69$
  (Wisotzki \etal\ 2002).  Panel (b) displays the best-fit S\'ersic
  model surface brightness profile, which is characterized by a
  S\'ersic index of $n=3.51\pm 0.04$, half-light radius $r_e=4.4\pm
  0.1$ kpc, and an axis ratio of $b/a=0.82\pm 0.01$. The best-fit
  morphological properties are consistent with those expected from an
  early-type galaxy.  Panel (c) displays the residual image after
  subtracting the QSO PSFs and the best-fit S\'ersic model.  Panel (d)
  displays an azimuthally-averaged surface brightness profile of the
  lensing galaxy in the F814W bandpass.  Each data point and the
  associated error are calculated from an elliptical annulus of 0.1"
  (2 pixels) in width.  The best-fit S\'ersic from the 2D analysis is
  shown as a black line.  We note that the deviant points at $>8$ kpc
  are due to large residuals from subtracting the QSO light.}
\label{Figure 5}
\end{figure*}

\begin{figure*} 
\begin{center}
\includegraphics[width=160mm]{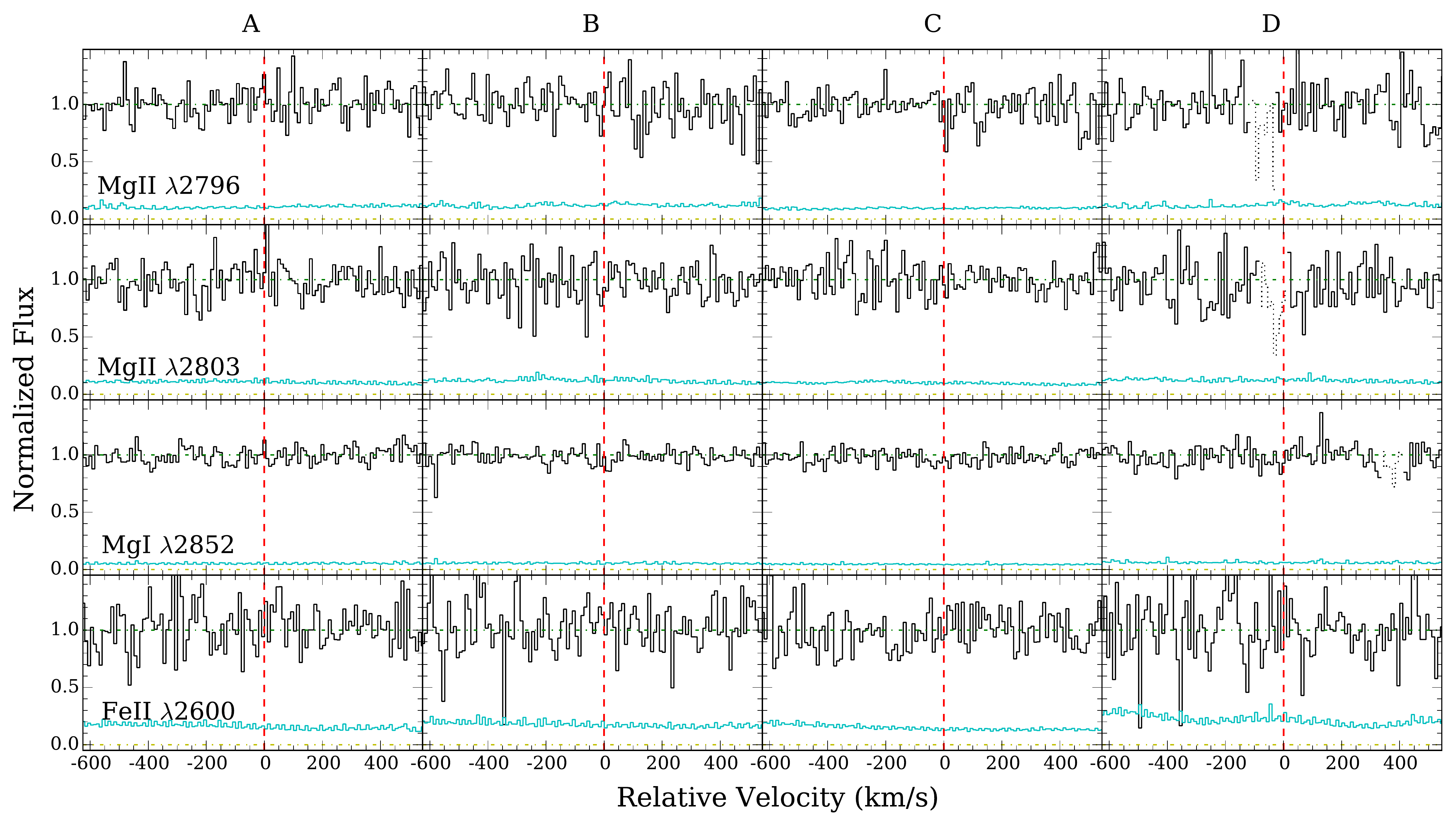}
\end{center}
\caption{Continuum normalized spectra along the lensed QSO sightlines
  HE\,0435$-$1223 $A$, $B$, $C$, $D$, at $d=7.5, 6.7, 7.5$, and 6.2
  kpc (or 1.7, 1.5, 1.7, $1.4\,r_e$), respectively.  Zero velocity
  corresponds to the systemic redshift of the lensing galaxy at
  $z=0.454$.  The 1-$\sigma$ error spectrum is included in thin, cyan
  curve above the zero flux level.  Contaminating features have been
  dotted out for clarity.  In stark contrast to the ultra-strong
  Mg\,II absorber detected along both sightlines of HE\,0047$-$1756
  (Figure 5), none of the four sightlines near the HE\,0435$-$1223
  lens reveals any cool gas.}
\label{Figure 6}
\end{figure*}

\subsection{HE\,0435$-$1223 Lens Galaxy at $z=0.454$}

%here
\begin{figure*} 
\begin{center}
\includegraphics[width=160mm]{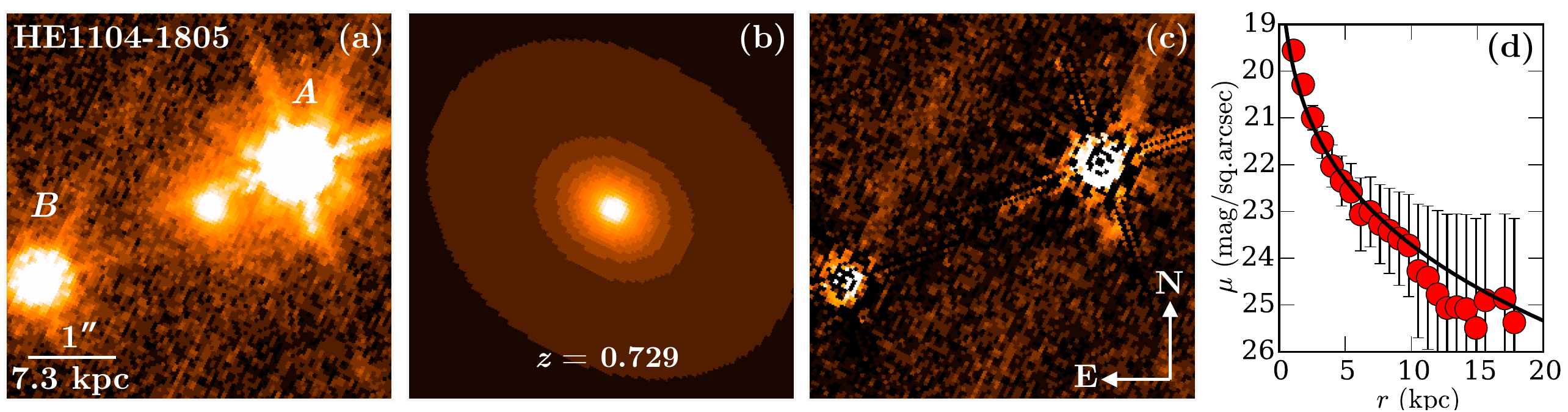}
\end{center}
\caption{(a) HST/WFPC2-PC F814W image of the field around
  HE\,1104$-$1805.  The lensing galaxy at $z=0.729$ is at the center
  of the panel surrounded by the doubly-lensed QSO at $z_{\rm
    QSO}=2.31$ (Wisotzki \etal\ 2000).  Panel (b) displays the
  best-fit S\'ersic model surface brightness profile, which is
  characterized by a S\'ersic index of $n=4.4\pm 0.2$, half-light
  radius $r_e=8.2\pm 0.2$ kpc, and an axis ratio of $b/a=0.77\pm
  0.01$. The best-fit morphological properties are consistent with
  those expected from an early-type galaxy.  Panel (c) displays the
  residual image after subtracting the QSO PSFs and the best-fit
  S\'ersic model.  Panel (d) displays an azimuthally-averaged surface
  brightness profile of the lensing galaxy in the F814W bandpass.  Each
  data point is calculated from an elliptical annulus Each data point
  and the associated error are calculated from an elliptical annulus
  of 0.1" (2.5 pixels) in width.  The best-fit S\'ersic from the 2D
  analysis is shown as a black line.  We note that the deviant points at
  $>10$ kpc are due to large residuals from subtracting the QSO
  light.}
\label{Figure 7}
\end{figure*}

The lensing galaxy of HE\,0435$-$1223 at $z=0.454$ was
spectroscopically identified by Morgan \etal\ (2005) and confirmed by
Eigenbrod \etal\ (2006).  This lens resides in a galaxy group with 12
spectroscopically confirmed members and a group velocity dispersion of
$\sigma_{\rm group}=520_{-80}^{+70}$ \kms\ (Wilson \etal, in
preparation).  We note, however, that galaxy G22 in Morgan et
al. (2005) at $4.4''$ from the lensing galaxy is spectroscopically
identified at $z=0.7818$ by Chen \etal\ (2014), and is therefore not
part of the group.  Our two-dimensional surface brightness profile
analysis has yielded a best-fit S\'ersic index of $n=3.51\pm 0.04$, a
half-light radius of $r_e=4.4$ kpc, and an axis ratio of $b/a=0.82\pm
0.01$ (Figure 5), consistent with measurements of Kochanek \etal\
(2006).  The best-fit morphological parameters characterize the lens
as an early-type galaxy, consistent with the old stellar population
($>1$ Gyr) and a lack of on-going star formation (${\rm
  SFR}<0.07\,{\rm M}_\odot\,{\rm yr}^{-1}$) inferred from both the
rest-frame UV continuum flux (Table 3) and the spectral indices
presented in Table 4.  Integrating the best-fit S\'ersic profile, we
find the total apparent magnitudes in the F555W, F814W, and F160W
bandpasses of $AB\rm{(F555W)}=21.30\pm0.04 $, $AB\rm{(F814W)}=
19.36\pm0.03$, and $AB\rm{(F160W)}=18.34\pm0.04$.  Our photometry for
this lens galaxy is in general agreement with the magnitudes reported
by Kochanek \etal\ (2006).  The observed apparent magnitudes translate
to a rest-frame $B$-band absolute magnitude of $M_B=-21.4$
(corresponding to $1.7\,L_*$ at $z=0.45$), a rest-frame optical color
of $g-r= 0.74$, and a total stellar mass of $\log\,M_*/M_\odot=11.1$
for the lensing galaxy.  The inferred dark matter halo mass is
$\log\,M_h/M_\odot=12.7-13.4$, consistent with
$\log\,M_h/M_\odot=13.3\pm 0.4$ from Kochanek \etal\ (2006) based on
the lensing properties of this system.  In summary, we find that the
lensing galaxy of HE\,0435$-$1223 is a quiescent super $L_*$ galaxy.

At $z=0.454$, the projected distances between the lensing galaxy and
four lensed QSO images are $d=7.5, 6.7, 7.5$, and 6.2 kpc or 1.7, 1.5,
1.7, $1.4\,r_e$ for images $A$, $B$, $C$, and $D$, respectively.  Similar to
the doubly lensed QSO HE\,0047$-$1756, the quadruply-lensed QSO
sightlines probe both the gaseous halo at small projected distances
and the interstellar medium of a massive, early-type galaxy.  Adopting
the same $L_x$-$L_B$ relation of O'Sullivan \etal\ (2001), we infer
$L_x\approx (2-3.5)\times 10^{41}\,{\rm erg}\,{\rm s}^{-1}$ for the
lensing galaxy.  

\begin{center}
\begin{table}
\scriptsize
\centering
    \caption{Constraints on the absorption properties around the HE\,0435$-$1223 Lens.}
    \begin{tabular}{ccccc}
      \hline
      & \multicolumn{1}{c}{$\log\,N({\rm Fe\,II})$} & \multicolumn{1}{c}{$\log\,N({\rm Mg\,II})$} & \multicolumn{1}{c}{$\log\,N({\rm Mg\,I})$} & \multicolumn{1}{c}{$\log\,N({\rm Ca\,II})$} \\
      \hline
      \hline
      $A$   &  $<12.5$  &   $<11.8$    &   $<11.0$  & $<11.5$ \\
      $B$   &  $<12.5$  &   $<11.9$    &   $<11.0$  & $<11.6$ \\
      $C$   &  $<12.4$  &   $<11.8$    &   $<10.9$  & $<11.5$ \\
      $D$   &  $<12.6$  &   $<12.0$    &   $<11.1$  & $<11.5$ \\
      \hline %\hline
    \label{HE0435 Table}
  \end{tabular}
\end{table}
\end{center}

\begin{figure}
\begin{center}
\includegraphics[width=77mm]{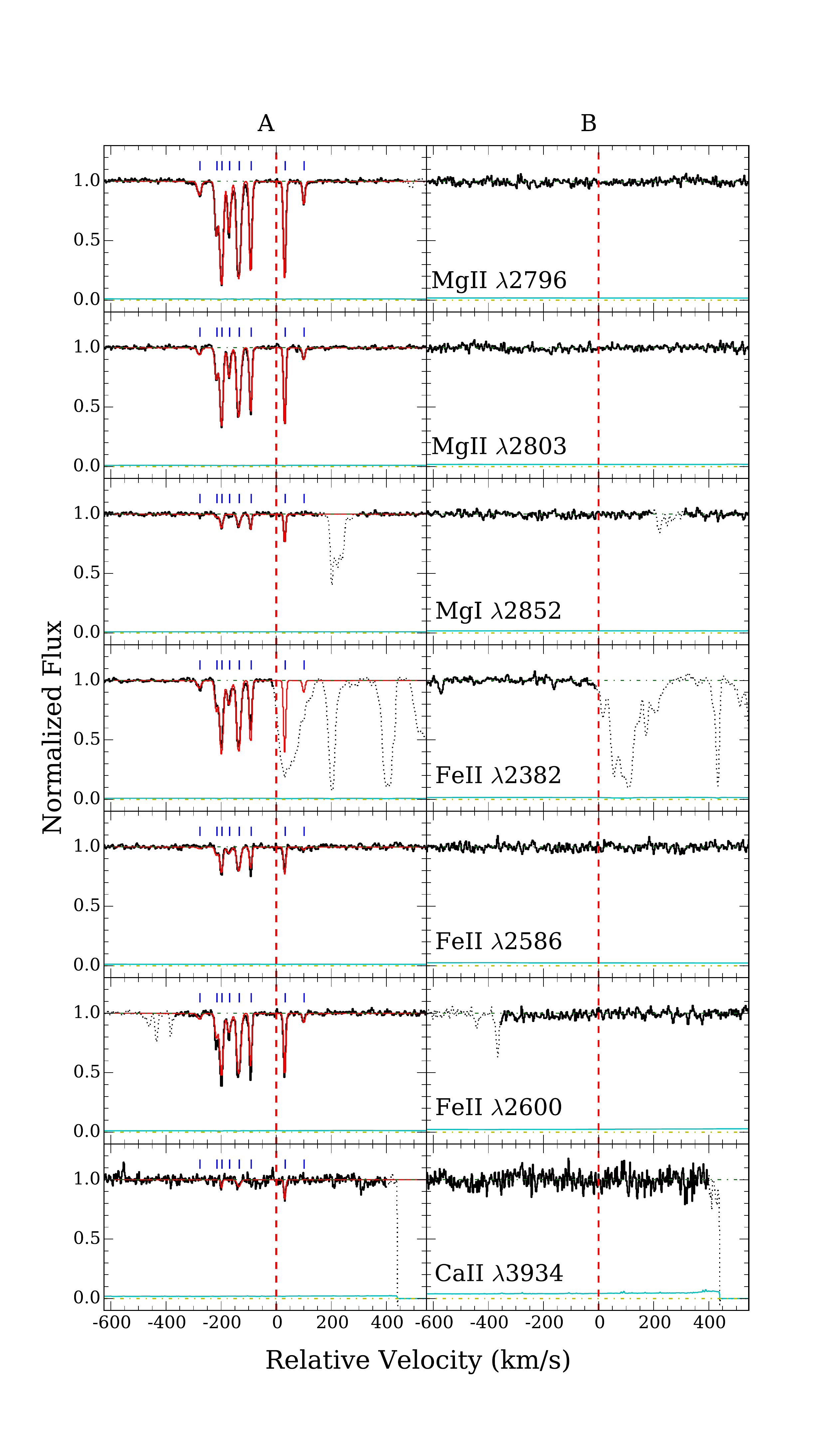}
\end{center}
\caption{Continuum normalized absorption profiles of different
  transitions along the lensed QSO sightlines $A$ (left) at $d=8.1$
  kpc (or $\approx r_e$) and $B$ (right) at $d=15.1$ kpc (or
  $1.8\,r_e$) from the HE\,1104$-$1805 lens.  Zero velocity
  corresponds to the systemic redshift of the lensing galaxy at
  $z=0.729$.  The 1-$\sigma$ error spectrum is included in thin, cyan
  curve above the zero flux level.  The blue tickmarks at the top of
  each left panel indicates the location of individual components
  included in the Voigt profile analysis (see \S\ 3.2), and the
  best-fit Voigt profile models are included in red.  The best-fit
  model parameters were found by a simultaneous $\chi^2$ analysis to
  the HIRES and UVES spectra. The HIRES and UVES spectra were combined
  for display purposes only, as shown here. Contaminating features
  have been dotted out for clarity.  Only sightline $A$ exhibits
  chemically-enriched cool gas characterized by six primary absorbing
  components that span $\approx 330$ \kms\ in line-of-sight velocity,
  while sightline $B$ exhibits no low-ionization transitions in the
  HIRES or UVES spectrum.}
\label{Figure 8}
\end{figure}

In contrast to the ultra-strong Mg\,II absorbers found in the
vicinities of the HE\,0047$-$1756 lens, no trace of ionic absorption
is detected near the HE\,0435$-$1223 to sensitive limits.  We measure
2-$\sigma$ upper limits of the underlying gas column densities for all
observed transitions based on the absorption equivalent width limits
presented in Table 5.  The column density constraints are presented in
Table 7.  In Figure 6, we present the continuum-normalized QSO spectra
within $\pm\,500$ \kms\ of the systemic redshift of the lensing
galaxy.  The strong limits afforded by the MIKE spectra indicates a
lack of cool gas in the inner halo of this massive, quiescent galaxy.

\subsection{HE\,1104$-$1805 Lens Galaxy at $z=0.729$} 

The lensing galaxy of HE\,1104$-$1805 at $z=0.729$ was
spectroscopically identified by Lidman \etal\ (2000).  Our
two-dimensional surface brightness profile analysis has yielded a
best-fit S\'ersic index of $n=4.4\pm 0.2$, a half-light radius of
$r_e=8.2$ kpc, and an axis ratio of $b/a=0.77\pm 0.01$ (Figure 7).
The best-fit morphological parameters indicate that the lens is also
an elliptical galaxy, consistent with a lack of on-going star
formation (${\rm SFR}<0.1\,{\rm M}_\odot\,{\rm yr}^{-1}$) inferred the
rest-frame UV continuum flux (Table 3).  Integrating the best-fit
S\'ersic profile, we estimate the total apparent magnitudes in the
F555W, F814W, and F160W bandpasses, and find
$AB\rm{(F555W)}=22.42\pm0.06 $, $AB\rm{(F814W)}= 20.04\pm0.04$, and
$AB\rm{(F160W)}=18.77\pm0.06$.  The observed apparent magnitudes
translate to a rest-frame $B$-band absolute magnitude of $M_B=-22.4$,
corresponding to $3.5\,L_*$ at $z=0.7$ according to Faber \etal\
(2007).  In addition, we find a rest-frame optical color of $g-r=
0.71$ and a total stellar mass of $\log\,M_*/M_\odot=11.4$ for the
lensing galaxy.  The inferred dark matter halo mass is
$\log\,M_h/M_\odot=13.3-14.3$.  In summary, we find that the lensing
galaxy of HE\,1104$-$1805 is a massive, quiescent galaxy.

At $z=0.729$, the projected distance between the lens and QSO image $A$
is $d_\mathrm{A}=8.1$ kpc or $\approx r_e$, and the projected distance
between the lens and QSO image $B$ is $d_\mathrm{B}=15.1$ kpc or
$1.8\,r_e$.  Similar to the previous two lenses, the two sightlines
probe both the gaseous halo at small projected distances and the
interstellar medium of an elliptical galaxy.  Adopting the $L_x$-$L_B$
relation of O'Sullivan \etal\ (2001), we infer $L_x\approx 3\times
10^{42}\,{\rm erg}\,{\rm s}^{-1}$ for the lens of HE\,1104$-$1805. 

For the two lensed QSO sightlines, we have high $S/N$ and high
spectral resolution echelle spectra, and interestingly the inner halo
of the HE\,1104$-$1805 lens exhibits absorption properties that lie
between the two previous cases.  Only sightline $A$ at $d=8.1$ kpc (or
$\approx r_e$) exhibits a modest amount of cool gas, as indicated by
the presence of a moderately strong Mg\,II absorber with
$W_r(2796)=0.6$ \AA\ (Smette \etal\ 1995).  Associated Fe\,II, Mg\,I,
and Ca\,II absorption transitions are also detected at this location
(Table 5).  Sightline $B$ at about twice the distance away, $d=15.1$ kpc
(or $1.8\,r_e$), does not show any trace of cool gas.  We place a
2-$\sigma$ upper limit of $W_r\rm{(2796)}=0.01$ \AA\ for possible
underlying cool gas (Table 5).

Similar to the Mg\,II absorbers found in the vicinities of the
HE\,0047$-$1756 lens, the strongest components (3 \& 5) along
HE\,1104$-$1805 $A$ may contain a non-negligible amount of H\,I with
$\log\,N({\rm H\,I})\apg 17$ while the remaining weaker components are
optically thin absorbers.  Unlike the the HE\,0047$-$1756 lens,
however, no edge-leading signature is found for the absorber near the
HE\,1104$-$1805 lens.  The absorption kinematics in the vicinities of
the HE\,1104$-$1805 lens is shown in Figure 8.  The moderately strong
Mg\,II absorber at $d=8.1$ kpc (or $1\,r_e$) is resolved into six
dominant components of comparable strength, spanning $\approx 330$
\kms\ in line-of-sight velocity.  The observed velocity spread along
sightline $A$ is comparable to the maximum circular velocity, $v_{\rm
  max}\approx 340$ \kms, expected for halos of $5\times
10^{12}\,M_\odot$.  Results from our Voigt profile analysis are
presented in Table 8. 

\begin{center}
\begin{table*}
\scriptsize
\centering
    \caption{Absorption properties around the HE\,1104$-$1805 Lens.}
    \begin{tabular}{ccrrcrcrcrcr}
      \hline
      & & & \multicolumn{1}{c}{$v_c$} & \multicolumn{2}{c}{Fe\,II} & \multicolumn{2}{c}{Mg\,II} & \multicolumn{2}{c}{Mg\,I} & \multicolumn{2}{c}{Ca\,II} \\
\cmidrule(lr){5-6} \cmidrule(lr){7-8} \cmidrule(lr){9-10} \cmidrule(lr){11-12}
      Sightline  & \multicolumn{1}{c}{$n_c$} & \multicolumn{1}{c}{component} & \multicolumn{1}{c}{(km/s)} &  $\log\,N_c$ & \multicolumn{1}{c}{$b_c$ (km/s)} &  $\log\,N_c$ & \multicolumn{1}{c}{$b_c$ (km/s)} &  $\log\,N_c$ & \multicolumn{1}{c}{$b_c$ (km/s)} &  $\log\,N_c$ & \multicolumn{1}{c}{$b_c$ (km/s)} \\
      \hline
      \hline
      A   &  8 &  1 &  $-277.7$ & $11.78\pm 0.06$ &  $8.9\pm 1.8$ & $11.69\pm 0.03$ & $8.9\pm 0.9$ &      $<10.2$     &     ...      & $<10.8$ & ... \\
          &    &  2 &  $-215.7$ & $12.34\pm 0.02$ &  $6.4\pm 0.4$ & $12.27\pm 0.01$ & $6.2\pm 0.3$ & $10.41\pm 0.12$ & $5.8\pm 2.9$ & $<10.7$ & ... \\
          &    &  3 &  $-197.3$ & $12.87\pm 0.01$ &  $6.7\pm 0.2$ & $12.82\pm 0.01$ & $6.3\pm 0.1$ & $11.08\pm 0.03$ & $6.2\pm 0.7$ & $10.96\pm 0.27$ & $1.0\pm 5.6$  \\
          &    &  4 &  $-169.8$ & $12.33\pm 0.02$ &  $8.9\pm 0.6$ & $12.25\pm 0.01$ & $6.7\pm 0.2$ &      $<10.1$     &     ...      & $<10.7$ & ... \\
          &    &  5 &  $-134.9$ & $12.92\pm 0.01$ &  $8.6\pm 0.2$ & $12.80\pm 0.01$ & $8.1\pm 0.1$ & $11.08\pm 0.03$ & $7.0\pm 0.7$ & $11.16\pm 0.09$ & $7.8\pm 2.4$ \\
          &    &  6 &   $-91.5$ & $12.66\pm 0.01$ &  $4.3\pm 0.2$ & $12.61\pm 0.01$ & $4.4\pm 0.1$ & $11.00\pm 0.02$ & $3.5\pm 0.6$ & $<10.7$ & ...  \\
          &    &  7 &   $+32.1$ & $12.75\pm 0.02$ &  $3.4\pm 0.3$ & $12.74\pm 0.01$ & $3.5\pm 0.1$ & $11.28\pm 0.01$ & $2.7\pm 0.4$ & $11.39\pm 0.04$ & $2.4\pm 1.2$  \\
          &    &  8 &  $+101.3$ & $11.80\pm 0.10$ &  $4.7\pm 2.2$ & $11.75\pm 0.02$ & $5.1\pm 0.5$ &      $<10.1$     &     ...      & $<10.7$ & ...  \\
      \hline
      B   &  ... &  ... &  $0.0$  &     $<11.0$    &  ... & $<10.8$  & ... & $<10.3$ & ...  & $<$11.0 & ... \\
      \hline %\hline
    \label{HE1104 Table}
  \end{tabular}
\end{table*}
\end{center}

\section[]{Analysis and Results}

\begin{figure} 
\begin{center}
\includegraphics[width=80mm]{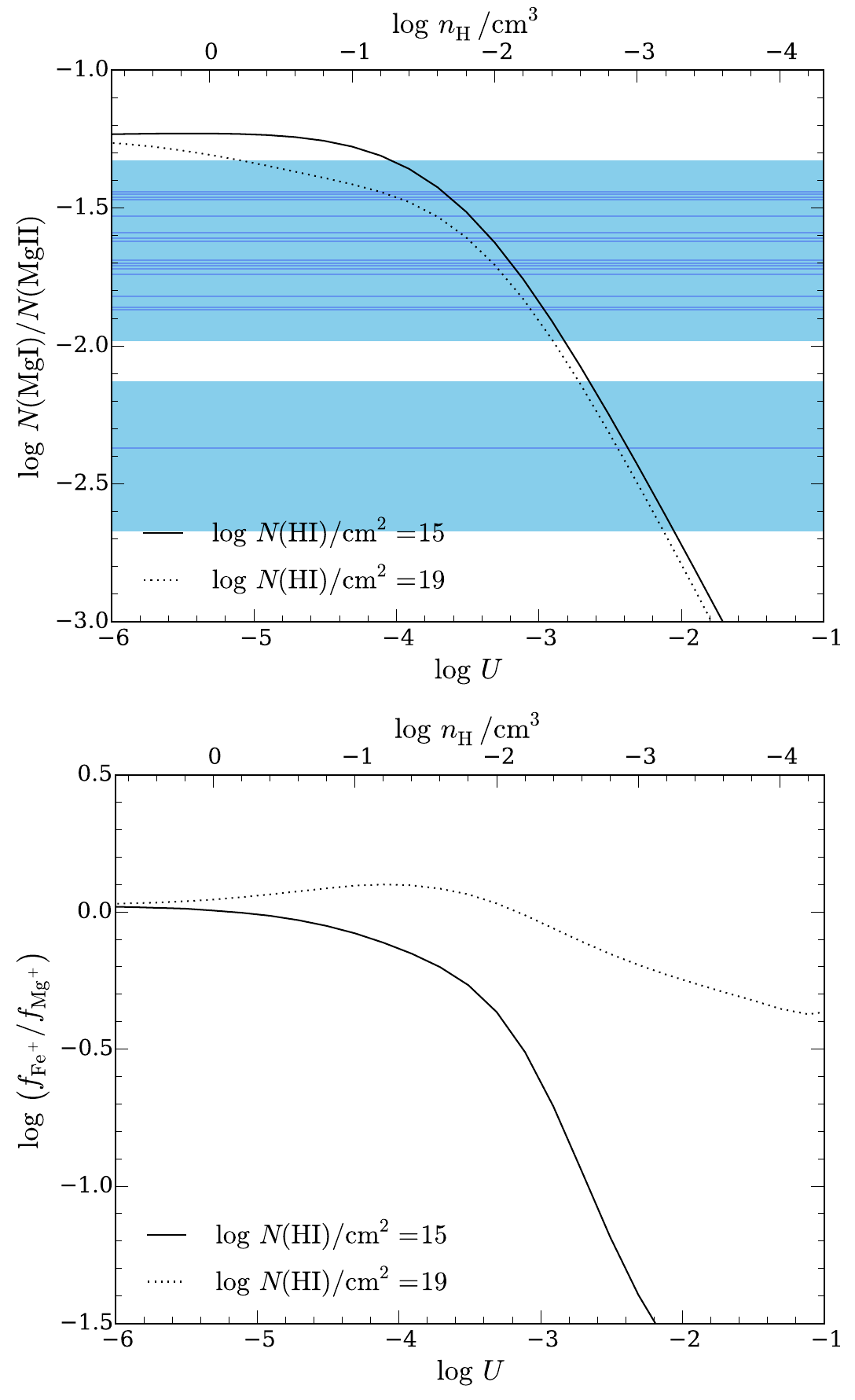}
\end{center}
\caption{Expectations from Cloudy photo-ionization calculations
  (Ferland \etal\ 2013; v.\ 13.03).  The {\it top} panel shows the
  expected $N({\rm Mg\,I})/N({\rm Mg\,II})$ versus $U$ for
  photo-ionized gas of $T=10^4$ K, 0.1 solar metallicity, and two
  different $N({\rm H\,I})$.  The observed Mg\,I to Mg\,II column
  density ratios (Tables 6 \& 8) are displayed in horizontal blue
  lines, with the band showing the 1-$\sigma$ measurement
  uncertainties.  Comparisons between observations and model
  predictions for both optically-thin ($\log\,N({\rm H\,I})=15$) and
  optically-thick ($\log\,N({\rm H\,I})=19$) gas constrain $U$ to be
  between $\log\,U\approx -3.9$ and $\log\,U\approx -2$ Experimenting
  with different gas metallicities (from 0.01 solar to solar) does not
  change the result.  The {\it bottom} panel shows the ionization
  fraction of Fe$^+$ relative to that of Mg$^+$ as a function of $U$.
  The model predictions show that independent of gas metallicity the
  observed $N({\rm Fe\,II})/N({\rm Mg\,II})$ ratio represents a
  conservative lower limit of the underlying $({\rm Fe}/{\rm Mg})$ for
  optically thin gas ($\log\,N({\rm H\,I})=15$), while in the
  optically thick regime the observed relative ionic abundance roughly
  reflects the underlying total $({\rm Fe}/{\rm Mg})$ ratio.}
\label{Figure 9}
\end{figure}

Our multi-sightline absorption-line search has revealed a diverse
range of halo gas properties at $d=1-2\,r_e$ from three lensing
galaxies at $z=0.4-0.7$.  While strong Mg\,II, Fe\,II, and Mg\,I
absorption features are found at the redshift of the double lens for
HE\,0047$-$1756 in both sightlines, these absorption features are
observed in only one of the two sightlines at the redshift the
HE\,1104$-$1805 lens and no absorbers are found in any of the four
sightlines near the lens for HE\,0435$-$1223.  Incidentally,
HE\,0435$-$1223 is the only one of the three lenses in our study known
to reside in a group environment (Wilson \etal, in preparation).  Here
we examine the physical properties (such as temperature and ionization
state), relative abundance pattern, and spatial coherence of gas
kinematics in the vicinities of the two lensing galaxies in
HE\,0047$-$1756 and HE\,1104$-$1805.

\subsection{Photo-ionized Cool Gas Associated with Lensing Galaxies}

All Mg\,II absorbers detected in our study have associated absorption
transitions due to Fe\,II\,$\lambda\,2600$, Fe\,II\,$\lambda\,2586$,
Mg\,I\,$\lambda\,2852$, and Ca\,II\,$\lambda\,3934$.  Comparisons of
the relative line widths and absorption strengths between these
transitions allow us to constrain both the temperature and ionization
state of the gas.  It is immediately clear from Tables 6 \& 8 that for
each component, different ionic transitions are found to share a
consistent best-fit Doppler parameter which is roughly $1-2$ times the
instrument resolution.  Recall from \S\ 3.2 that our Voigt profile
analysis was carried out with the velocity offsets of individual
components fixed across all transitions, but letting $\log\,N_c$ and
$b_c$ vary freely for different transitions.  Fixing the $b_c$ value
of each component for all transitions would lead to $<0.1$ dex
differences in the best-fit column density.  The consistent best-fit
$b_c$ between Mg$^0$ and Mg$^+$ states lends strong support for the
hypothesis that these different ions originate in the same gaseous
clouds.  In addition, the consistent $b_c$ between Fe$^+$ and Mg$^+$,
which differ in mass by a factor of two, indicates that the line
broadening is driven by non-thermal motions and confirms the
expectation that Mg\,II absorbing gas is typically cool with
temperature $T\apl {\rm a\ few} \times 10^4$ K (e.g., Bergeron \&
Stasi\'nska 1986).

Given the relatively cool gas temperature, we consider the scenario in
which the gas is being photo-ionized and determine the ionization
state based on the observed relative abundances between Mg$^0$ and
Mg$^+$ ions.  Under the photo-ionization scenario, a key factor is the
ionization parameter $U$, which is defined as the number of ionizing
photons $\phi$ relative to the total hydrogen number density $n_{\rm
  H}$, $U\equiv \phi/c\,n_{\rm H}$.  For a fixed radiation field,
lower gas densities lead to higher $U$ parameters, and the gas is
expected to be more highly ionized.  Conversely, higher gas density
lead to lower $U$ parameters, and the gas is more neutral.  We perform
a series of photo-ionization calculations using the Cloudy package
(Ferland \etal\ 2013; v.\ 13.03) and construct a grid of models that
span a range in $U$, from $\log\,U=-6$ to $\log\,U=-1$ and a range in
gas metallicity, from $1/100$ solar to solar.  We consider both
optically-thin gas with $\log\,N({\rm H\,I})=15$ and optically-thick
gas with $\log\,N({\rm H\,I})=19$, which are representative of the
expected H\,I column density range for the absorbing clouds found near
the lensing galaxies.  For each photo-ionization model, we assume a
plane-parallel geometry for the absorbing gas of $T=10^4$ K, which is
illuminated on both sides by an updated version of the Haardt \& Madau
(2001) ionizing radiation field (HM05 in Cloudy) at $z=0.5$.  Then we
compute the expected relative abundance ratios between Mg$^0$, Mg$^+$,
and Fe$^+$ for gas that follows a solar abundance pattern.  While no
empirical knowledge is available for the neutral hydrogen
column density of the Mg\,II absorbers found in the vicinities of the
lensing galaxies, the photo-ionization models can be constrained based
on the observed column density ratio between Mg$^0$ and Mg$^+$.
Finally, we note that the Cloudy models we constructed are dust-free, 
although we will discuss the effects of dust depletion in \S\ 5.2.

The top panel of Figure 9 shows the expected $N({\rm Mg\,I})/N({\rm
  Mg\,II})$ versus $U$ for photo-ionized gas of $1/10$ solar
metallicty and different $\log\,N({\rm H\,I})$ (solid and dotted
curves), in comparison to obserations from Tables 6 \& 8 (blue lines
and associated bands to indicate the 1-$\sigma$ uncertainties.  
The observed Mg\,I to Mg\,II column density ratios constrain the
ionization parameter in the range from $\log\,U\approx -3.9$ to
$\log\,U\approx -2$ or gas density from $n_{\rm H}\apl 0.04\,{\rm
  cm}^{-3}$ to $n_{\rm H}\apl 4\times 10^{-4}\,{\rm cm}^{-3}$.  The
results are insensitive to the adopted gas metallicity.  The allowed
$U$ range is consistent with previous findings for weak Mg\,II
absorbers of $W_r(2796)<0.3$ \AA\ (e.g., Rigby \etal\ 2002) or for
Lyman limit systems at similar redshifts (e.g., Lehner \etal\ 2013).

\begin{figure} 
\begin{center}
\includegraphics[width=80mm]{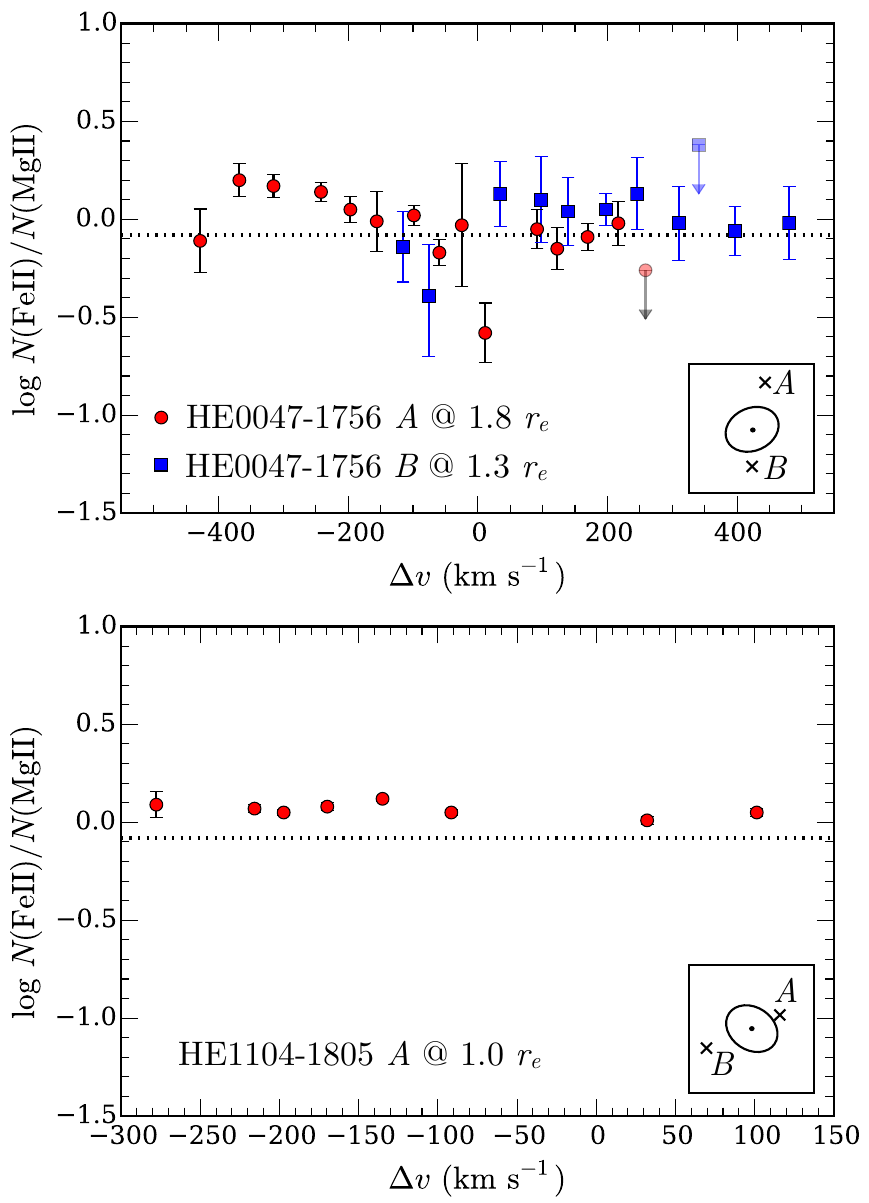}
\end{center}
\caption{The observed column density ratio of Fe$^+$ to Mg$^+$ versus
  velocity offset, $\Delta\,v$, for individual components along both
  sightlines $A$ and $B$ of HE\,0047$-$1745 in the {\it top} panel and
  along sightline $A$ of HE\,1104$-$1805 in the {\it bottom} panel.
  Error bars associated with individual data points represent the
  1-$\sigma$ uncertainties, with upper limits indicating an absence of
  Fe\,II absorption.  The dotted line indicates the expected solar
  (Fe/Mg) ratio of $\log\,({\rm Fe}/{\rm Mg})=-0.08$ (Asplund 2006) for
  visual comparisons.  The inset shows the geometric alignment between
  the lensing galaxy shown as the ellipse and the lensed QSO images.
  Both the axis ratio and position angle of the ellipse are
  representative of the observed morphological properties of the
  lens.}
\label{Figure 10}
\end{figure}

\subsection{Uniform Super-solar (Fe/Mg) Across $\apg 400$ \kms\ at $d\approx 1-2\,r_e$}

The observed column density ratio between Fe$^+$ and Mg$^+$ ions
allows us to estimate the underlying total elemental abundance ratio
between iron and magnesium, $({\rm Fe}/{\rm Mg})$, even though we
cannot constrain the gas metallicity due to unknown $N({\rm H\,I})$
for these absorbers.  Constraining the relative abundance between iron
and magnesium is particularly interesting, because both core collapse
and Type Ia supernovae (SNe) contribute to the observed iron abundance
(e.g., Tsujimoto \etal\ 1995) while magnesium is an $\alpha$ element
generated primarily in massive stars and core-collapse SNe (e.g.,
Nomoto \etal\ 2006).  Specifically, every Type Ia supernova is
expected to release $\sim 0.7\,M_\odot$ of iron, while at the same
time producing $\apl 0.02\,M_\odot$ of magnesium (e.g., Thielemann
\etal\ 1986; Iwamoto \etal\ 1999).  The relative [Fe/Mg] ratio
therefore provides a quantitative measure of the relative contribution
of massive stars to the chemical enrichment in galaxies (e.g.,
Tsujimoto \etal\ 1995; Ferreras \& Silk 2002; de Plaa \etal\ 2007).

Furthermore, Mg$^+$ and Fe$^+$ share similar ionization potentials (15
eV and 16.2 eV, respectively) and are the dominant ionization states of
the respective species in both neutral and cool photo-ionized medium.
Observations of $N({\rm Fe\,II})/N({\rm Mg\,II})$ should reflect
closely the intrinsic total elemental abundance ratio, $({\rm Fe}/{\rm
  Mg})$.  Specifically,
\begin{equation}
\log\,\left(\frac{{\rm Fe}}{{\rm Mg}}\right)=\log\,\frac{N({\rm Fe\,II})}{N({\rm Mg\,II})} - \log\,\frac{f_{{\rm Fe}^+}}{f_{{\rm Mg}^+}},
\end{equation}
where $f_{{\rm Fe}^+}$ is the fraction of Fe in singly ionized state
and $f_{{\rm Mg}^+}$ is the fraction of Mg in singly ionized state.
The bottom panel of Figure 9 shows the ratio of the ionization
fractions for the Fe$^+$ and Mg$^+$ states as a function of $U$.
Adopting the model predictions for photo-ionized gas, we expect that
the observed $N({\rm Fe\,II})/N({\rm Mg\,II})$ ratio represents a
conservative lower limit of the underlying $({\rm Fe}/{\rm Mg})$ for
optically-thin gas, which is likely the case for weak components of
$\log\,N({\rm Mg\,II})<14$ near the lensing galaxies (Tables 6 \& 8).
In the optically-thick regime, which is likely the case for the strong
components (e.g., component 2 along HE\,0047$-$1756 $B$), the observed
$N({\rm Fe\,II})/N({\rm Mg\,II})$ reflects the underlying total $({\rm
  Fe}/{\rm Mg})$ ratio.

We present the observed $N({\rm Fe\,II})/N({\rm Mg\,II})$ versus
velocity offset, $\Delta\,v$, for individual components in Figure 10.
Observations for both sightlines $A$ and $B$ of HE\,0047$-$1745 are
presented in the {\it top} panel and observations for sightline $A$ of
HE\,1104$-$1805 in the {\it bottom} panel.  Error bars associated with
individual data points represent the 1-$\sigma$ uncertainties, with
upper limits indicating absence of Fe\,II absorption.

Figure 10 displays two striking features.  First, the observed $N({\rm
  Fe\,II})/N({\rm Mg\,II})$ ratios are high, exceeding the typical
solar abundance pattern.  Second, the dispersion in the observed
$N({\rm Fe\,II})/N({\rm Mg\,II})$ ratio is small among different
components across the full velocity range.  The only exceptions are
components 10 \& 15 along HE\,0047$-$1756 $A$ and possibly component 2
along HE\,0047$-$1756 $B$.  Excluding these outliers, we find a median
of $\langle\,\log\,N({\rm Fe\,II})/N({\rm Mg\,II})\,\rangle \approx 0$
and dispersion of 0.11 and 0.09 dex, respectively, for sightlines $A$
and $B$ near HE\,0047$-$1756.  For HE\,1104$-$1805 $A$, we find a
median of $\langle\,\log\,N({\rm Fe\,II})/N({\rm Mg\,II})\,\rangle =
0.06$ and dispersion of 0.03 dex.  

Such homogeneity in abundance ratio across multiple components is not
commonly seen in absorbers uncovered along random sightlines.  For
example, weak Mg\,II absorbers often display a component-to-component
$N({\rm Fe\,II})/N({\rm Mg\,II})$ ratio that varies by $\approx 0.3$
dex to more than 0.4 dex over $\Delta\,v\approx 50$ \kms\ (e.g.,
Narayanan \etal\ 2008) and high-redshift DLAs display up to 0.9 dex
difference between components separated by $\Delta\,v\apl 100$
\kms\ (e.g., Fox \etal\ 2014).  In contrast, the scatter of the
observed $N({\rm Fe\,II})/N({\rm Mg\,II})$ in the vicinities of the
two massive lensing galaxies is $<0.1$ dex across $\Delta\,v> 400$
\kms.

\begin{figure*} 
\begin{center}
\includegraphics[width=140mm]{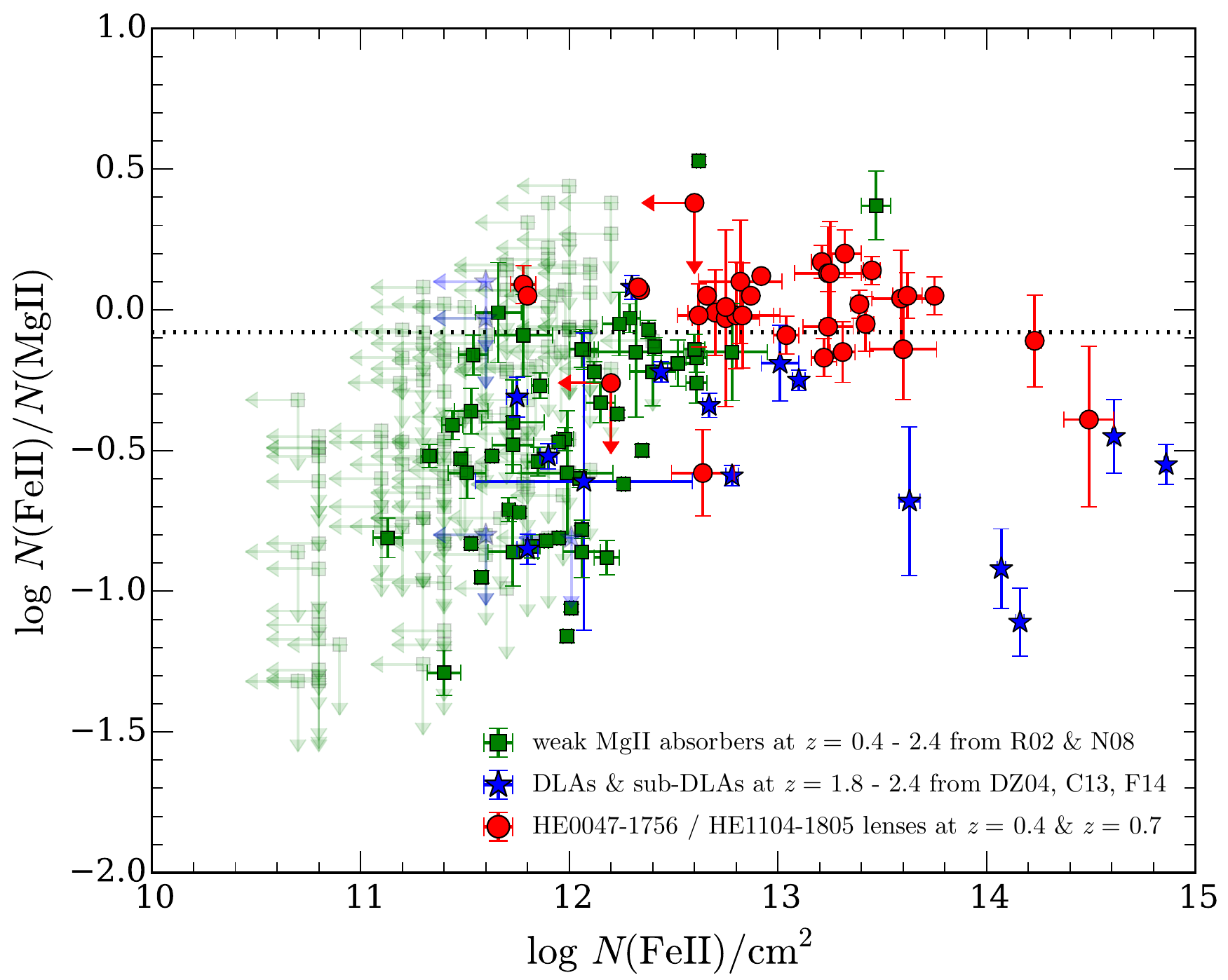}
\end{center}
\caption{Observed Fe\,II to Mg\,II column density ratios versus
  $N({\rm Fe\,II})$ both for absorbers selected from random sightlines
  and for halo gas around lensing galaxies from our study.  Each data
  point represents a single absorbing component resolved in
  high-resolution QSO spectra.  Our observations near lensing galaxies
  are presented in red, circles with error bars showing the
  corresponding measurement uncertainties.  Measurements from
  high-redshift ($z=1.8-2.4$) DLAs and sub-DLAs from
  Dessauges-Zavadsky (2004), Crighton \etal\ (2013), and Fox
  \etal\ (2014) are shown in blue star symbols.  Measurements for weak
  Mg\,II absorbers of $W_r(2796)<0.3$ \AA\ at $0.4<z<2.4$ from Rigby
  \etal\ (2002) and Narayanan \etal\ (2008) are shown in green
  squares.  Absorbing components with no associated Fe\,II absorption
  transitions are shown as downward and left-pointing arrows with the
  data point indicating the 2-$\sigma$ upper limit in $N({\rm
    Fe\,II})$.  We have greyed out weak Mg\,II data points without
  associated Fe\,II detections for clarity.  Like Figure 10, the
  dotted line indicates the solar (Fe/Mg) abundance pattern for visual
  calibrations.  With the exception of three outlying points, halo
  clouds around lensing galaxies occupy a unique $\log\,N({\rm
    Fe\,II})/N({\rm Mg\,II})$--$\log\,N({\rm Fe\,II})$ space,
  separated from either the weak Mg\,II absorber group or
  DLAs/sub-DLAs.  The scatter in $\log\,N({\rm Fe\,II})/N({\rm
    Mg\,II})$ is small, $\sim 0.1$ dex, in the halos around lensing
  galaxies.  Including necessary ionization fraction corrections, our
  observations imply a super solar Fe to Mg abundance pattern for the halo
  gas around these lensing galaxies.}
\label{Figure 11}
\end{figure*}

For comparisons, we include in Figure 11 the observed $N({\rm
  Fe\,II})/N({\rm Mg\,II})$ versus $N({\rm Fe\,II})$ for individual
components found in weak Mg\,II absorbers of $W_r(2796)<0.3$ \AA\ at
$z=0.4-2.4$ from Rigby \etal\ (2002) and Narayanan \etal\ (2008), and
in DLAs or sub-DLAs at $z=1.8-2.4$ from Dessauges-Zavadsky (2004),
Crighton \etal\ (2013), and Fox \etal\ (2014), together with the
observations for individual components found near lensing galaxies.
It is immediately clear that the relatively high $N({\rm
  Fe\,II})/N({\rm Mg\,II})$ ratios are not common among QSO
absorption-line systems found along random sightlines.  While weak
Mg\,II absorbers have on average lower $N({\rm Fe\,II})$ per component
and exhibit a significant scatter in the observed $N({\rm
  Fe\,II})/N({\rm Mg\,II})$ ratio, only a few components have been
found to show enhancement in $N({\rm Fe\,II})$ relative to $N({\rm
  Mg\,II})$.  DLAs and sub-DLAs, which are thought to originate in
young star-forming galaxies, have on average higher $N({\rm Fe\,II})$
per component and exhibit a significant $\alpha$-element enhancement
abundance pattern.  Neither one of these absorber populations occupies
the same parameter space as the Fe-rich components found near the
lensing galaxies.  The observed discrepancy strongly suggests a
different origin for the gas found along the sightlines near the two
lensing galaxies.

From the Cloudy photo-ionization models presented in the bottom panel
of Figure 9, we find $\log\,(f_{{\rm Fe}^+}/f_{{\rm Mg}^+})\apl -0.3$
for the allowed range of $U$ in the optically thin regime and
$-0.3\apl \log\,(f_{{\rm Fe}^+}/f_{{\rm Mg}^+})\apl 0$ in the
optically thick regime.  Applying these conditions to Equation (3)
leads to 
\begin{equation}
\log\,({\rm Fe}/{\rm Mg})>\log\,N({\rm Fe\,II})/N({\rm  Mg\,II}).
\end{equation} 
We therefore conclude that the observed $N({\rm Fe\,II})/N({\rm
  Mg\,II})$ represents a conservative minimum for the underlying total
elemental abundance ratio, $({\rm Fe}/{\rm Mg})$, particularly in
relatively weak components of $\log\,N({\rm Mg\,II})<14$, for which
the gas is likely optically thin to the ionizing radiation field.
Adopting a solar abundance pattern of $\log\,({\rm Fe}/{\rm
  Mg})_\odot=-0.08$ (Asplund \etal\ 2006), we estimate a super solar Fe
to Mg abundance pattern of $[{\rm Fe}/{\rm Mg}]\equiv \log\,({\rm
  Fe}/{\rm Mg}) - \log\,({\rm Fe}/{\rm Mg})_\odot\apg 0.1$ for
individual absorbing components across a velocity range of $\approx
400$ \kms.

We note that the presence of Ca\,II absorption (Tables 6 \& 8) also
strongly implies the presence of dust (e.g., Savage \& Sembach 1996;
Wakker \& Mathias 2000), which would further alter the observed
elemental abudance pattern if not accounted for.  Observations of
local interstellar clouds have shown that Fe is more depleted in
cool phase by as much as 0.5 dex relative to Mg (Savage \& Sembach
1996; Lauroesch \etal\ 1996).  If differential dust depletion is
significant, then the inferred $[{\rm Fe}/{\rm Mg}]$ will be even
higher.

\subsection{Probing Spatial Coherence of Gas Kinematics Across $d\approx 8$ kpc of the HE\,0047$-$1756 Lens}

A primary goal of the multi-sightline study is to examine the spatial
coherence of chemically-enriched gas across the halo (e.g., Chen
\etal\ 2014).  In our current study, the doubly lensed HE\,0047$-$1756
exhibits an ultra-strong Mg\,II absorber along both lensed QSO
sightlines separated by $\approx 8$ kpc in projected distance, and
offers a unique opportunity to examine possible correlation in the gas
kinematics at $d\approx 1.5\,r_e$ on the opposite sides of the
lensing galaxy.  We have noted two remarkable similarities in the
absorption signatures between the two sightlines: (1) an edge-leading
signature with the strongest absorption component moving at the
highest blueshifted velocity and (2) strong correlation between a
large number of components over a velocity range of $\approx 420$
\kms\ along each sightline (see \S\ 4.1).  These observed kinematic
signatures suggest strong coherence on scales of 8 kpc in the gas
motion across the inner halo of the HE\,0047$-$1756 lens galaxy,
similar to the strong coherence found on scales of 10 kpc at $d\approx
40-50$ kpc from a blue, $L_*$ galaxy at $z=0.4188$ by Chen
\etal\ (2014).  A natural explanation of the edge-leading signature is
a rotating disk (e.g., Lanzetta \& Bowen 1992).  However, if the two
sightlines probe the opposite sides of a rotating disk, then the
edge-leading feature of the second sightline is expected to be flipped
from the first (e.g., Prochaska \& Wolfe 1997).  This is not what is
observed.  Instead, the edge-leading profiles are merely offset by
$\Delta\,v\approx 350$ \kms\ between the two sightlines.  It is not
clear a gaseous stream connecting to the lensing galaxy would produce
such consistent edge-leading profiles on the opposite sides of the
galaxy.

Further insights into the connection between the observed absorbing
gas along two closely separated sightlines may be gained from
comparing the chemical abundance pattern.  In \S\ 5.3 and in Figure
10, we show that while the majority of the absorbing components in
HE\,0047$-$1756 exhibit a uniform super solar $[{\rm Fe}/{\rm Mg}]$
relative abundance with a small scatter, three components have
non-negligible likelihood of being Mg-rich ($\alpha$-element
enhanced).  These are components 10 \& 15 along sightline $A$ and
component 2 along sightline $B$.  We note that component 2 in
sightline $B$ is especially interesting, given the observed large gas
column density that exceeds all other components found near this lens.
The large error bar in the Fe/Mg ratio indicates that there is a
non-negligible probability that this gas may be Mg-rich.  Because
Mg-rich gas does not share the same SNe~Ia enhanced chemical
enrichment history as Fe-rich gas, these possible Mg-rich components
could provide additional clues for the nature of the gas along these
two sightlines.

Using the observed Fe/Mg ratio to identify absorption components
sharing a common origin, we reproduce the best-fit absorption profiles
of Mg\,II\,$\lambda$\,2796, Fe\,II\,$\lambda$\,2600, and
Mg\,I\,$\lambda$\,2852 transitions from Figure 4 in the bottom left
panel of Figure 12 but highlight the few possibly Mg-rich components
in shaded green.  Excluding possible Mg-rich components, we find that
the correlation between the absorption profiles along sightlines $A$
and $B$ weakens, with no clear maximum between $\Delta\,v=0$ and
$\Delta\,v=600$ \kms, due to a diminishing edge-leading signature
along sightline $B$.  We therefore caution that line-of-sight
kinematics alone may not be sufficient to determine the spatial
coherence of gas at multiple locations.

\section[]{Discussion}

The widely dispersed line-of-sight gas kinematics around the lens of
HE\,0047$-$1756 is especially remarkable when compared to the H\,I gas
detected in nearby elliptical/S0 galaxies (e.g., Oosterloo
\etal\ 2007; Serra \etal\ 2012).  Roughly 40\% of nearby elliptical/S0
galaxies exhibit extended H\,I gas out to $\sim 30$ kpc in radius with
increasing velocity shear up to $\pm\,200$ \kms\ at the edges of the
observed H\,I structures (e.g., Serra \etal\ 2012).  The observed
line-of-sight velocity spread around the HE\,0047$-$1756 lens would
still exceed the largest velocity dispersion ($\approx 250$ \kms)
known at the peak H\,I emission near the center of local elliptical
galaxies (see Figure 6 of Serra \etal\ 2012).  Alternatively, the
observations of the HE\,0047$-$1756 lens can be explained if these
gas-rich nearby ellipticals are surrounded by high-dispersion low
column density clouds that fall below the typical column density limit
of $N({\rm H\,I})\approx 5\times 10^{19}$ \cmjj\ in local 21 cm
observations.

The observed relative (Fe/Mg) abundance pattern offers new clues for
understanding the origin of chemically-enriched cool gas in massive
quiescent halos.  In particular, strong metal-line absorbers are
commonly attributed to starburst driven outflows (e.g., Murray
\etal\ 2011; Booth \etal\ 2013).  However, it is difficult to explain
the presence of metal-enriched cool gas near quiescent galaxies based
on the outflow scenario.  Additional explanations that have been
proposed include stripped satellites due to tidal interactions or ram
pressure force (e.g., Wang 1993; Agertz \etal\ 2009; Gauthier 2013),
gas accreted from the IGM (e.g., Rauch \etal\ 1997; Nelson
\etal\ 2013), as well as in-situ cloud formation from thermal
instabilities (e.g., Mo \& Miralda-Escude 1996; Maller \& Bullock
2004; Sharma \etal\ 2012).
Different scenarios would predict different chemical abundance patterns.
For example, if the gas is pre-enriched by early-generation star
formation (which is expected for newly accreted IGM as well as ISM in
and or stripped from blue satellites), then it is also expected to be
$\alpha$-element enriched (e.g., Rauch \etal\ 1997).  On the other
hand, if the gas has significant contribution from SNe Ia ejecta
(which is expected for ISM in or stripped from red satellites), then
it is expected to show Fe enhancement.

For the cool, Fe-rich gas, the observed super solar (Fe/Mg) ratio
indicates a significant contribution to the chemical enrichment from
SNe~Ia.  For reference, the expected SNe Ia contribution relative to
total (Type Ia and core-collapse SNe combined) in the solar abundance
pattern is $f_{\rm Ia}\approx 15$\% (e.g., Tsujimoto \etal\ 1995; de
Plaa \etal\ 2007).  Here we estimate $f_{\rm Ia}$ for the observed
Fe-rich gas around massive lensing galaxies based on the observed
$N({\rm Fe\,II})/N({\rm Mg\,II})$, which shows a median of
$\langle\,\log\,N({\rm Fe\,II})/N({\rm Mg\,II})\,\rangle \approx
0-0.06$ (\S\ 5.2). From Equation (4), these values represent
conservative lower bounds for $\log\,({\rm Fe}/{\rm Mg})$.  Adopting
the expected yields for Type Ia and core-collapse SNe from Iwamoto
\etal\ (1999), we estimate a minimum fractional contribution of SNe Ia
to the chemical enrichment of $f_{\rm Ia}\approx 20$\%.  The estimated
SNe~Ia contribution around massive, lensing galaxies is comparable to
what is found for intracluster medium (e.g., de Plaa \etal\ 2007).
Furthermore, SNe~Ia occur in low-mass, evolved stars (e.g., Maoz
\etal\ 2014) with long lifetimes, which also implies that the gas has
been enriched to relatively high metallicities by previous generations
of massive stars.  Indeed, observations of stellar atmospheres find an
increasing Fe abundance with increasing metallicity (e.g., McWilliam
1997).  We therefore expect the Fe-rich halo gas to also have a high
metallicity.  This expectation is at least consistent with what is
found for the few Fe-rich, weak Mg\,II absorbers by Rigby \etal\
(2002).

Based on the uniform super solar (Fe/Mg) ratios with a small
dispersion and the expected high metallicity, we argue that the
Fe-rich absorbing components (which dominate the total absorption
width) originate in SNe~Ia enriched inner regions at radius $r\sim d$
from the lensing galaxies.  The large velocity spread ($\Delta\,v\apg
400$ \kms) can be attributed to recent SNe~Ia ejecta (Figure 12).  In
addition to chemically enriching their environment, SNe~Ia ejecta can
interact with the surrounding gas and deposit thermal energy that
increases the velocity dispersion of the gas.  We consider stellar
winds from low-mass stars an unlikely source for the observed Fe-rich
gas, because observations of nearby elliptical galaxies have uncovered
a predominantly $\alpha$-enhanced abundance pattern in these stars
(e.g., Kuntschner \etal\ 2010).

A SNe~Ia origin is qualitatively consistent with previous finding that
the radial distribution of Type Ia SNe in early-type galaxies is
consistent with the S\'ersic profile describing the stellar light of
the galaxies (F\"{o}rster \& Schawinski 2008).  Recall also that the
lensed QSO sightlines that have revealed this cool, Fe-rich gas pass
through the interstellar space of the lensing galaxies at merely
$1-1.8\,r_e$.

Alternatively, the observed kinematic and chemical signatures may be
explained by Fe-rich gas expelled from red satellites.  Large-scale
surveys of nearby galaxies have shown that massive, red galaxies have
a significantly higher fraction of red satellites (e.g., Prescott
\etal\ 2011).  Star formation in these red satellites is shut off as a
result of gas removal upon entering the host halo (e.g., Larson
\etal\ 1980; Kawata \& Mulchaey 2008).  As the remaining stars
continue to evolve in the satellites, we also expect increased
contribution from SNe~Ia to the chemical enrichment in the surrounding
gas, resulting in increasingly Fe-rich abundance patterns.

The absence of absorption features along sightline $B$ at $1.8\,r_e$
from the lens for HE\,1104$-$1805 and in all four sightlines at
$1.4-1.7\,r_e$ from the lens for HE\,0435$-$1223 is qualitatively
consistent with the partial gas covering fraction expected from either
stripped satellites or clumpy SNe Ia ejecta.  However, additional
factors from the galactic environments may have also played an
important role in depleting the cool gas near the quad-lens of
HE\,0435$-$1223.  This lensing galaxy is the only one of the three
lenses in our study known to reside in a group environment (e.g.,
Morgan \etal\ 2005; Wilson et al., in preparation) and the only one
that appears to be devoid of cool gas at small projected distances to
the lens.  A similar example is found by Johnson \etal\ (2014), who
reported a transparent sightline at $d<20$ kpc from a pair of
interacting galaxies.  Interactions between group members may have
heated or tidally stripped cool gas from the inner halo of the lens
galaxy.  Follow-up studies of the galaxy environments of the other two
fields will cast important insights into possible environmental role
in the CGM properties of these lenses.

It has been shown that chemically-enriched, cool gas is not uncommon
in massive, quiescent halos (see e.g., Gauthier \etal\ 2009, 2010;
Bowen \& Chelouche 2011; Zhu \etal\ 2014).  The mean gas covering
fraction is found to be $\kappa_{\rm Mg\,II}> 15$\% at $d\apl 100$ kpc
(Gauthier \& Chen 2011; Huang \etal\ 2016).  However, the physical
nature of such gas remains unknown.  Based on the suppressed velocity
dispersion between Mg\,II gas and galaxies, and the observed
preferential geometric alignment of Mg\,II absorbers with filaments,
Huang \etal\ (2016) have recently argued that the observed Mg\,II
absorbers are best-explained by a combination of cool clouds formed in
thermally unstable hot halos and satellite accretion through
filaments.  We expect that Mg\,II gas produced in these processes
should exhibit an abundance pattern that reflects an $\alpha$-element
enhancement.

\begin{figure*} 
\begin{center}
\includegraphics[width=170mm]{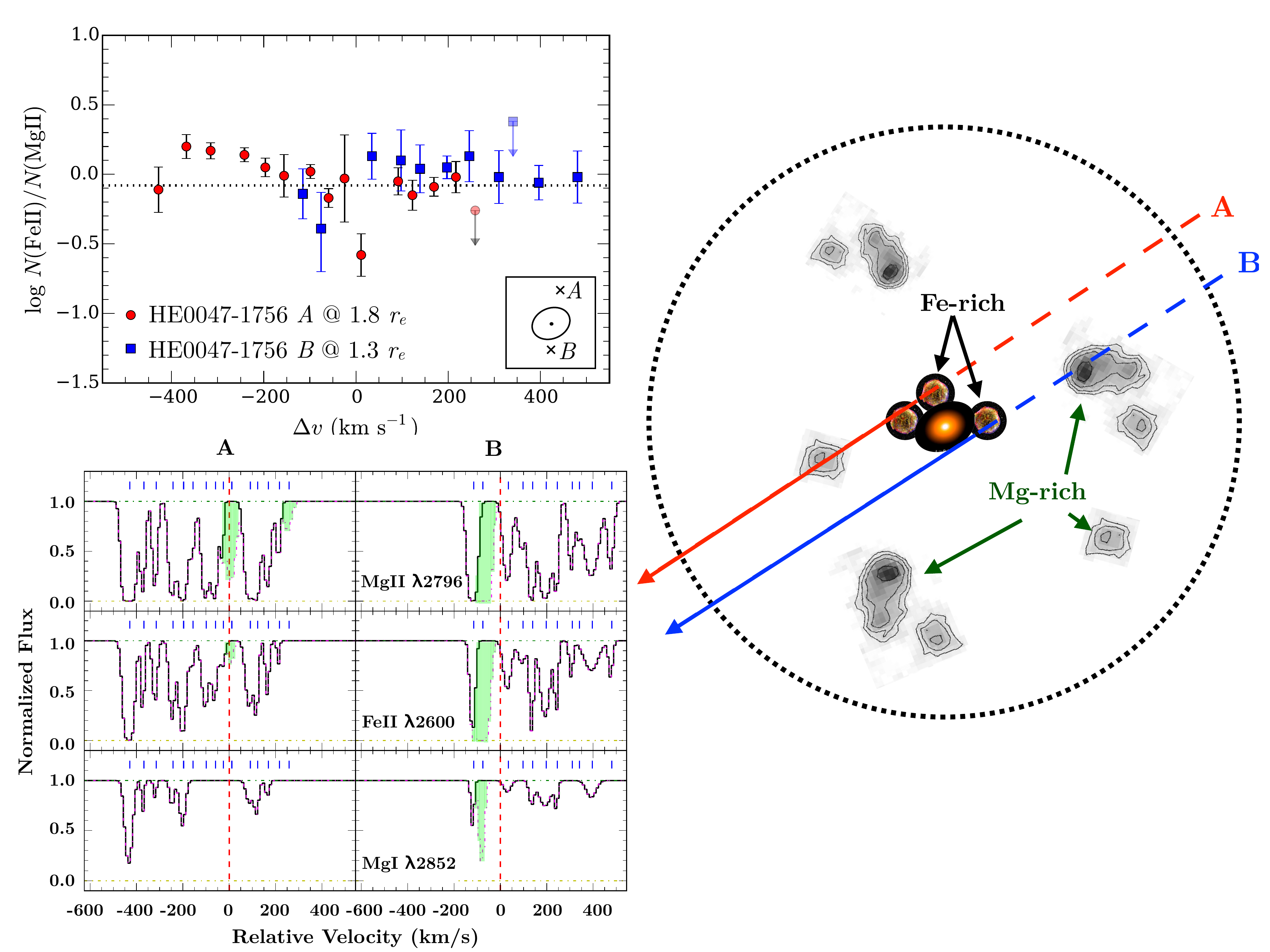}
\end{center}
\caption{Cartoon illustrating the possibility of combining the
  observed $N({\rm Fe\,II})/N({\rm Mg\,II})$ of individual components
  (upper-left panel) and their kinematic signatures (lower-left panel)
  for identifying the origins of the absorbing clouds in the diagram
  on the right.  We adopt observations for the HE\,0047$-$1756 lens
  for the illustration.  In the absorption profiles panel, we present
  the best-fit model absorption profiles from Figure 4 for
  Mg\,II\,$\lambda$\,2796, Fe\,II\,$\lambda$\,2600, and
  Mg\,I\,$\lambda$\,2852 transitions, but with possible Mg-rich
  components dotted out and shaded in pale-green.  The resemblance in
  the kinematic signatures of Fe-rich absorbing gas between sightlines
  $A$ and $B$ is weakened.  Because of the required SNe~Ia
  contributions to the elevated Fe enrichment, as well as the
  remarkably uniform (Fe/Mg) relative abundances for these Fe-rich
  absorbing components, we propose that these components originate in
  the inner regions at radius $r\sim d$, with the observed large
  velocity spread driven by recent SNe~Ia ejecta (the schematic
  diagram on the right).  While the large velocity shear ($\approx
  350$ \kms) between the Fe-rich gas across two sightlines may be
  explained by bulk rotation, we consider this scenario unlikely
  because the edge-leading features along sightlines on the opposite
  sides of the rotating disk are expected to be flipped from each
  other (Lanzetta \& Bowen 1992; Prochaska \& Wolfe 1997), contrary to
  what is observed here.
We highlight the possibility of Mg\,II-rich components originating in
stripped gas from blue satellites or condensed halo clouds at larger
distances ($r\gg d$) that were enriched by early episodes of star
formation.}
\label{Figure 12}
\end{figure*}

Interestingly, the strongest component (component 2) at
$\Delta\,v\approx -75$ \kms\ in sightline $B$ and a moderately strong
component (component 10) at $\Delta\,v\approx 11$ \kms\ in sightline
$A$ of the HE\,0047$-$1756 lens are consistent with an
$\alpha$-element enhancement, although we cannot rule out Fe-rich gas
as the origin of component 2 in sightline $B$.  We note that component
2 toward HE\,0047$-$1756 $B$ contains a significant amount of neutral
gas with $\log\,N({\rm H\,I})>19$.  The inferred large H\,I column
density resembles the high velocity clouds (HVCs) found in the Milky
Way Halo or DLAs and sub-DLAs found along random QSO sightlines.  Both
HVCs and DLAs exhibit $\alpha$-element enhancement driven by core
collapse SNe (e.g., Richter 2006; Wolfe \etal\ 2005), and both are
primarily found in/near star-forming galaxies.  This is, however,
inconsistent with the old stellar population found for the lensing
galaxy with no trace of a cool interstellar medium.  Incidentally, a
blue galaxy is seen at $\approx 1.6"$ southwest of the lens.  At
$z=0.408$, this galaxy would be at $d\approx 5.4$ kpc from sightline
$B$ and nearly twice the distance from sightline $A$ which shows
weaker Mg-rich components.  It is possible that this gas, if Mg-rich,
originates in an infalling gas cloud or stripped gas from a blue
satellite as it orbits around the lensing galaxy (see the cartoon
illustration in Figure 12).

While it is also possible that this possibly Mg-rich gas arises in the
ISM or halo gas of a faint satellite that is missed in the glare of
the QSO light, we consider this an unlikely scenario.  Both Gauthier
\etal\ (2010) and Huang \etal\ (2016) have estimated possible
contributions from the CGM of satellite galaxies to the observed
Mg\,II absorption around massive luminous red galaxies (LRGs),
assuming that the satellites can retain their gas content while moving
in the hot halos of LRGs. These authors found that the expected
contribution from the CGM of satellites is $<20\,(10)$\% at
$d<50\,(100)$ kpc.  Considering only ISM, however, which is roughly 10
times smaller in size than the CGM, we expect a contribution of
$<2\,(1)$\% at $d< 50\,(100)$ kpc.  Because the cross section of blue
satellites is negligible, the observed Mg-rich gas is unlikely to
originate in an unresolved blue satellite.  Follow-up spectroscopy to
measure the redshift and star-forming properties of the blue object
will provide necessary data for further studies.

In summary, if the observed ISM gas of massive galaxies is locally
enriched by SNe~Ia and and the halo is more dominated by pre-enriched
gas from early-generation of young stars, then a radial dependence in
[Fe/Mg] should be observed, with sightlines at small projected
distances showing a higher fraction of high [Fe/Mg] absorbers.  We
plan to test this hypothesis using a larger sample of luminous red
galaxy and QSO pairs.

\section[]{Conclusions} 

We have carried out multi-sightline absorption spectroscopy of cool
gas around three lensing galaxies at $z=0.4-0.7$, HE\,0047$-$1756,
HE0435$-$1223, and HE\,1104$-$1805.  The fields are selected from
wide-separation gravitational lens systems with angular separations
$\theta\apg 1.5''$, which ensure that each lensed QSO serves as an
independent probe of the foreground lensing galaxy.  All three lensing
galaxies are massive with estimated total stellar masses of
$\log\,M_*/M_\odot=10.9-11.4$.  They exhibit spectral and photometric
properties that are characteristic of nearby elliptical galaxies with
half-light radii of $r_e=2.6-8$ kpc.  

For each lensing galaxy, the lensed QSO sightlines occur at projected
distances of $1-2\,r_e$, providing for the first time an opportunity
to probe spatial variations of the cool gas content both in the
interstellar space at $r\sim r_e$ and in the halos at larger radii
$r\gg r_e$ of quiescent galaxies.  The main results from our
absorption-line analysis of halo gas properties at the redshift of the
lensing galaxies are summarized below.

(1) We observe a diverse range of absorbing gas properties among the three
lensing galaxies.  Specifically, both sightlines at $1.8\,r_e$ and
$1.3\,r_e$ on the opposite sides of the lens for HE\,0047$-$1756
exhibit an ultra-strong Mg\,II absorber at the redshift of the lens.
The lensing galaxy of HE\,1104$-$1805 exhibits a moderately strong
Mg\,II absorber along sightline $A$ at $\approx r_e$, but no trace of
absorption within a sensitive upper limit along sightline $B$ at
$1.8\,r_e$.  The strong absorbers found near the two lensing galaxies
are resolved into $8-15$ narrow components with line-of-sight velocity
spread of $\Delta\,v\approx 300-600$ \kms.  In contrast, none of the
four sightlines at $d=1.4-1.7\,r_e$ from the quadruple lens for
HE\,0435$-$1223 exhibits detectable Mg\,II absorption, and this is the
only one of the three lenses in our study known to reside in a group environment.

(2) The strong Mg\,II absorbers with associated Fe\,II, Mg\,I, and
Ca\,II absorption at $d=1-2\,r_e$ reveal chemically-enriched cool gas
($T\apl\,{\rm a\ few}\,\times\,10^4$ K) around massive, evolved
galaxies at intermediate redshifts.  The observed large column
densities for Mg$^+$, Fe$^+$, and Ca$^+$ suggest a significant neutral
gas fraction, similar to Lyman limit clouds.  The two strongest
components may even be DLAs.

(3) The majority of the absorbing components exhibit a super solar
Fe/Mg ratio with only three components displaying a likely Mg-rich
($\alpha$-element enhancement) abundance pattern.  All three possible
Mg-rich components occur at the redshift of the lens for
HE\,0047$-$1756.  The Fe-rich components, along both sightlines of the
lens for HE\,0047$-$1756 and one sightline of the lens for
HE\,1104$-$1805, show a uniform super solar $({\rm Fe}/{\rm Mg})$
ratio with a dispersion of $<0.1$ dex, across the full velocity range
of the absorbers ($\Delta\,v\approx 300-600$ \kms).

(4) The absorbers uncovered on the opposite sides of the lens for
HE\,0047$-$1756 appear to share a common edge-leading absorption
signature, suggesting a spatial coherence on scales of $\apg 8$ kpc.
However, the spatial coherence is diminished when excluding possible
Mg-rich gas.  The kinematic signatures of the remaining components
share little resemblance between the two sightlines.

Given the predominantly old stellar populations in the lensing
galaxies and the observed small scatter in Fe/Mg, we argue that the
Fe-rich gas (which dominates the total absorption width) originates in
the inner regions (at radius $r\sim 1-2\,r_e$) of the lensing
galaxies, with the observed large velocity spread driven by recent
SNe~Ia ejecta.  Stellar winds from low-mass stars are an unlikely
source for the observed Fe-rich gas, because observations of nearby
elliptical galaxies have uncovered a predominantly $\alpha$-enhanced
abundance pattern in these stars.

We consider the possibility that the few possible Mg-rich components arise
in either stripped gas or pre-enriched halo gas at larger distances 
$r\gg d$, where the chemical enrichment is likely driven by young stars.  
Incidentally, a candidate gas-rich companion is seen in deep HST images of 
HE\,0047$-$1756.  Follow-up studies of the galaxy environment will provide 
further insights into the physical origin of these components.

In summary, we show that additional spatial constraints in
line-of-sight velocity and relative abundance ratios afforded by a
multi-sightline approach provide a powerful tool to resolve the
origin of chemically-enriched cool gas in massive halos.  Our study of
three lensing galaxies uncovers a broad range of absorbing gas
properties around massive, evolved galaxies.  A larger sample of
lensing galaxies is clearly necessary to determine whether the super
solar Fe/Mg abundance pattern is representative of cool gas at small
projected distances from evolved galaxies.  At the same time, single
QSO probes of foreground luminous red galaxies over a wide range of
projected distances should yield important constraints for the radial
dependence of [Fe/Mg], which will provide further insights into the
chemical enrichment history in massive halos.

\section*{Acknowledgments}

We thank an anonymous referee for thoughtful comments that helped
improve the presentation of the paper, Frederic Courbin for kindly
providing the spectra of the lensing galaxies of HE\,0047$-$1756 and
HE\,0435$-$1223 (data obtained using the ESO-VLT Unit Telescope 2
Kueyen under Programs 074.A-0563 and 075.A-0377), and Sebasti\'an
L\'opez for providing the echelle spectra of HE\,1104$-$1805 (data
obtained using the ESO-VLT Unit Telescope 2 Kueyen under Programs
067.A-0278 and 070.A-0439).  FSZ and HWC thank Sean Johnson for
important discussions and comments that helped improve the
presentation of the paper.  We also thank Rebecca Pierce for helpful
comments on an earlier draft of the paper.  MR thanks Andy McWilliam
for useful discussions and the National Science Foundation for support
through grant AST-1108815.  MLW and AIZ acknowledge funding from NSF
grant AST-1211874 and NASA grant ADP-10AE88G.  MLW also thanks the
Technology and Research Initiative Fund (TRIF) Imaging Fellowship
program for its support. In addition, HWC acknowledges the Aspen
Center for Physics, which is supported by National Science Foundation
grant PHY-1066293, and the organizers of the workshop on the ``physics
of accretion and feedback in the circumgalactic medium'' for a
productive visit in June 2015, during which components of the work
presented were accomplished.

\bsp

\label{lastpage}

\end{document}